\documentclass[11pt]{article}

\pdfoutput=1

\usepackage{amsmath,amssymb,amsfonts,amscd,mathrsfs,latexsym,amstext}
\usepackage{xcolor}
\usepackage[vcentermath]{youngtab}
\usepackage{tabu}
\definecolor{darkblue}{rgb}{0.1,0.1,.7}
\usepackage[colorlinks, linkcolor=darkblue, citecolor=darkblue, urlcolor=darkblue, linktocpage]{hyperref} 
\usepackage[square, comma, compress,numbers]{natbib}
\usepackage[]{graphicx}
\usepackage{geometry}
\usepackage[margin=10pt,font=small,labelfont=bf]{caption}
\usepackage{ifthen}
\usepackage{tikz}
\usepackage{subcaption}
\usepackage{booktabs,multirow}
\usepackage{hhline}
\usepackage{tablefootnote}
\usepackage{dsfont} 
\usepackage{braket}
\usepackage{bbold}
\usepackage{MnSymbol}
\usepackage{array}

\usepackage{titlesec}
\titleformat*{\section}{\large\bfseries}
\titleformat*{\subsection}{\normalsize\bfseries}
\titleformat*{\subsubsection}{\normalsize\it}
\titleformat*{\paragraph}{\normalsize\bfseries}
\titleformat*{\subparagraph}{\normalsize\bfseries}

\def \pd {\partial}
\def \bpd {\bar \partial}
\def \bq {\bar q}
\def \bz {\bar z}
\def \bp {\bar p}
\def \bJ {\bar J}
\def \calL {\mathcal L}
\def \calO {\mathcal O}
\def \calV {\mathcal V}
\def \calD {\mathcal D}
\def \bcalV {\overline{\mathcal V}}
\def \veps{\varepsilon}
\def \eps{\epsilon}
\def \id {\mathbb{1}}
\def \vphi{\varphi}
\def \bvphi{\bar \varphi}
\newcommand{\beq}{\begin{equation}}
\newcommand{\eeq}{\end{equation}}
\newcommand{\beqg}{\begin{equation} \begin{gathered}}
\newcommand{\eeqg}{\end{gathered} \end{equation}}
\newcommand{\beqa}{\begin{equation} \begin{aligned}}
\newcommand{\eeqa}{\end{aligned} \end{equation}}

\newcommand{\vect}[1]{\boldsymbol{#1}}

\def\bZ {\mathbb{Z}}
\def\l {\left}
\def\r {\right}

\newcolumntype{C}{>{$}c<{$}} %

\usepackage{accents}
\newlength{\dhatheight}

\numberwithin{equation}{section}
\usepackage[tocgraduated]{tocstyle}

\begin{document}

\vspace*{-.6in} \thispagestyle{empty}
\begin{flushright}
\end{flushright}
\vspace{1cm} {\large
\begin{center}
{\bf Two-dimensional $O(n)$ models and logarithmic CFTs}
\end{center}}
\vspace{1cm}
\begin{center}
{\bf Victor Gorbenko$^{a,b}$, Bernardo Zan$^{c,d}$}\\[2cm] 
{
$^{a}$  Institute for Advanced Study, Princeton, NJ 08540, USA\\
$^b$  Stanford Institute for Theoretical Physics, Stanford University, Stanford, CA 94305, USA}\\
$^c$ Department of Physics, Princeton University, Princeton, NJ 08544, USA\\
$^d$ Theoretical Particle Physics Laboratory (LPTP), Institute of Physics, EPFL, Lausanne, Switzerland
\vspace{1cm}%

\vspace{1cm}%
\end{center}

\vspace{4mm}

\begin{abstract}
We study $O(n)$-symmetric two-dimensional conformal field theories (CFTs) for a continuous range of $n$ below two. These CFTs describe the fixed point behavior of self-avoiding loops. There is a pair of known fixed points connected by an RG flow. When $n$ is equal to two, which corresponds to the Kosterlitz-Thouless critical theory, the fixed points collide. We find that for $n$ generic these CFTs are logarithmic and contain negative norm states; in particular, the $O(n)$ currents belong to a staggered logarithmic multiplet. Using a conformal bootstrap approach we trace how the negative norm states decouple at $n=2$, restoring unitarity. The IR fixed point possesses a local relevant operator, singlet under all known global symmetries of the CFT, and, nevertheless, it can be reached by an RG flow without tuning. Besides, we observe logarithmic correlators in the closely related Potts model.

\end{abstract}
\vspace{.2in}
\vspace{.3in}
\hspace{0.7cm} May 2020

\newpage

\setcounter{tocdepth}{3}

{
	\tableofcontents
}

\section{Introduction}
The critical $O(n)$ model is one of the best studied examples of fixed points both in condensed matter and high energy physics, and yet it keeps supplying us with new ideas. Our own interest in these theories is triggered by the fact that, for some range  of $n$, they give rise to a family of strongly coupled Conformal Field Theories (CFTs), and as such they provide an example of a line of fixed points parametrized by $n$.
In this work we will focus on the two-dimensional case, where such `conformal window' spans the range $-2\leq n\leq 2$. Interesting Renormalization Group (RG) phenomena tend to happen in such situations, which are expected to be quite generic and independent of the details of the microscopic theory.

 The physics outside the conformal window is also interesting in its own right. In particular, slightly `above' the window, for $n\gtrsim 2$, one expects to find complex fixed points and walking RG behavior, see \cite{Gorbenko:2018_1,Gorbenko:2018_2} and references therein for the detailed discussion. In this paper, however, we will mostly study the theory in the conformal window, with special interest for its upper end, $n\lesssim 2$. It turns out that a detailed study of this regime also reveals a plethora of curious RG effects. For example, we will find that the two-dimensional $O(n)$ model is a logarithmic CFT at  any point inside the conformal window, while for positive integer values of $n$ logarithmic multiplets recombine in a cumbersome way in order to form a unitary non-logarithmic (ordinary) subsector. In case of $n=1$  this subsector is just the critical Ising model, while for $n=2$ it is the Kosterlitz-Thouless fixed point \cite{Berezinsky:1970fr,Kosterlitz:1973xp}. Before we dive into the technical discussion of these effects, let us start with a brief introduction into what is already known about the $O(n)$ model. 

First of all, let us remind the reader what is meant by the continuous-$n$ $O(n)$ model. Indeed, for a person familiar with this model as a system of $O(n)$ spins placed on some lattice, or maybe a theory of $O(n)$-symmetric scalar fields, our claims may already appear strange.
 Let's start by a definition of the model as a spin system which works for integer $n$. Consider a maximum degree three lattice (for example a honeycomb lattice), %
with a spin $\vect{S}_i$, which is a $n$ dimensional vector of unit norm,  on every site coupled with the following Hamiltonian:
\beq
\beta H = -\sum_{\braket{i,j}}\log \left( 1+K \vect{S}_{i} \cdot \vect{S}_{j} \right) \label{eq:H2d}\,,
\eeq
where $\braket{i,j}$ indicates nearest neighbor sites, and $K$ is some coupling constant.

The spin partition function is
\beq
Z_{spins}=\int \left( \prod_x d \vect{S}_x\right)  e^{-\beta H}=\int \left( \prod_x d \vect{S}_x\right) \prod_{\braket{i,j}}\left( 1+K \vect{S}_i \cdot \vect{S}_j\right) 
\eeq
and, by expanding the right hand side and doing the integral, it can be written as a loop model (this was originally done in \cite{Domany}; see for example \cite{Jacobsen2012} or section 7.4.6 of \cite{DiFrancesco_book} for more explanations)

\beq
Z_{loops}=\sum_{loops}K^{N_{links}}n^{N_{loops}}\,,
\label{eq:Zloops}
\eeq
where the sum goes over all possible configurations of self and mutually avoiding closed loops on the lattice, $N_{links}$ is the total number of links in all loops, and $N_{loops}$ is the number of loops in a configuration. 
Here $n$ is just a parameter which can take any real value, in particular, the case $n=0$, corresponds to self-avoiding random walks \cite{deGennes}. 
Moreover, as was understood only very recently, so-defined loop model possesses a continuous-$n$ $O(n)$ well-defined categorical symmetry \cite{Binder:2019zqc}, so it indeed can, with full honesty, be called an $O(n)$ model for any $n$. Importantly, $n$ doesn't run with the RG \cite{Binder:2019zqc} and hence is a {\it parameter} of the theory, not a {\it coupling}, and choice of lattice is not important for the presence of $O(n)$ symmetry.

\begin{figure}
	\centering
	\includegraphics[scale=.3]{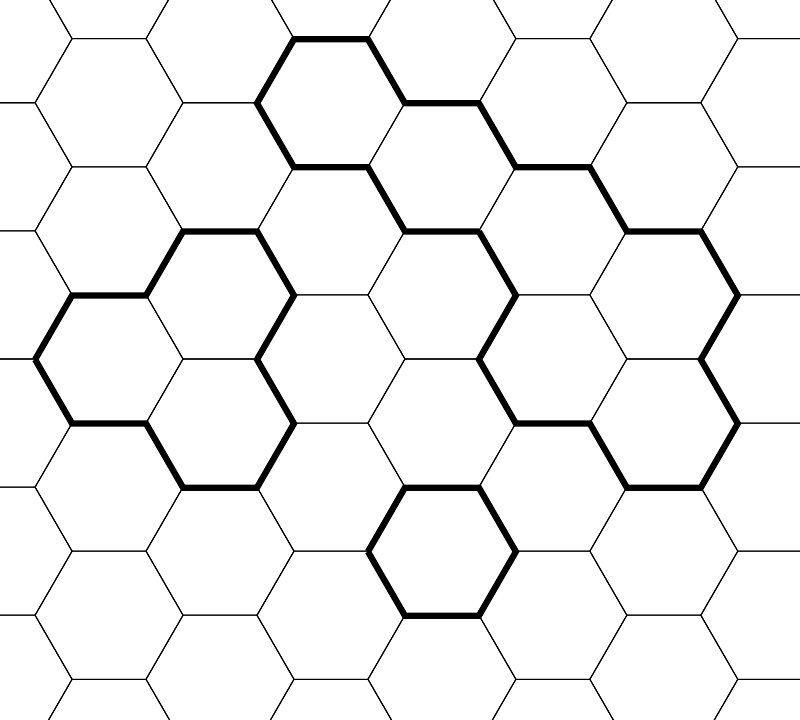}
	\caption{A loop configuration on the honeycomb lattice. In this example we have $N_{links}=40$ and $N_{loops}=3$.}
\end{figure}

We can recover the spin formulation operators in the loop formulation by introducing points at which a line of links can end, or by imposing some other geometrical constraints on the loops \cite{Nienhuis1987,cardy1994geometrical}. To give an example, the probability of spins $\vect{S}_x$ and $\vect{S}_y$ being correlated in the spin formulation is related to the expectation value of a line of links starting at  point $x$ and ending in $y$ in the loop formulation. These lattice constructions are very interesting, but they will not play a big role in our current investigation and we proceed to review the continuum limit of these loop models.

It is well known that in 2d for some range of $n$ and for a tuned value of $K$,\footnote{The critical value of $K$ on a honeycomb lattice is known exactly \cite{Nienhuis1982} {and rigorously \cite{Smirnov}} to be $K_c=\left(2+\sqrt{2-n} \right) ^{-1/2}$.} \eqref{eq:Zloops} has a continuum limit which gives rise to a CFT. 
The 2d loop model is equivalent to a 
solid-on-solid model, which, in turn, at criticality renormalizes to a gaussian scalar field \cite{diFrancesco:1987qf}. Thus, for example, the torus partition function of the theory, and hence the operator content, is known exactly for any $n$ in the conformal window. This analysis (as well as Monte-Carlo simulations \cite{Blote1989}) reveals that in fact there are actually two $O(n)$ fixed points, usually called critical and low-temperature (low-T) fixed points.\footnote{
The loop model we describe here has another fixed point, namely the $T=0$ fixed point described by fully packed loops \cite{FLP1,FLP2,Kondev_1996}. We will not consider this fixed point in our analysis. } As was first described in \cite{PhysRevLett.45.499}, at the upper end of the conformal window, $n=2$, they merge and annihilate, in agreement with the general fixed-point annihilation picture advocated in \cite{Kaplan:2009kr, Gorbenko:2018_1}. The above-described lattice construction assures that this picture is not just a formal analytic continuation in $n$, but for any $n$ the $O(n)$ CFT arises as an IR limit of a loop model. Note that degree three lattices also exist in 3 dimensions and corresponding loop models can also be constructed \cite{3dON}. This can be used to justify analytic continuation in $n$ also of the 3d critical $O(n)$ model.\footnote{The more complicated case of the 3d cubic lattice can be found in \cite{Chayes}.}

There is also an RG flow from the critical to the low-temperature phase, meaning that if we start from the critical point and we lower the temperature by a bit, we will flow to a non-trivial CFT in the IR, rather than a gapped phase. We will study this flow when it becomes perturbative, for $n \lesssim 2$. Harmless as it seems, this RG flow is quite unusual. In fact we will find that an IR fixed point contains a relevant singlet operator, so one would expect that in order to approach it some relevant singlet should be tuned to zero in the UV fixed point. However, the only relevant operator in the critical $O(n)$ model is the energy, which we do not tune. Instead the singlet which is relevant in the IR is irrelevant in the UV.\footnote{Sometimes such operators are called ``dangerously irrelevant'', although this term has several other meanings as well.} This phenomena was discovered in the context of loop models in \cite{Jacobsen:2002wu}, where it was pointed out that the IR phase of the $O(n)$ models depends on the presence of loop crossings. On the honeycomb lattice loop crossings are absent and the low-T fixed point is achieved without tuning. In spite of this, we do not find any additional symmetry in the CFTs that could correspond to the absence of crossing. In more phenomenological terms, there is a hierarchy problem in the $d=2$ $O(n)$ model, which is resolved without tuning. We will discuss this puzzle further in section \ref{sec:Dangerous}.

In the bulk of this paper we study the critical fixed point for a generic value of $n$, but with the aim of taking the $n\to2$ limit. The theory at generic $n$ is known to be non-unitary and contains an infinite set of primary operators, so unlike minimal models it is not solved exactly, even though some quantities, like the dimensions of operators, are known analytically for any $n$. It turns out, however, that even the two-point functions of primary operators in this theory are not immediately determined from this spectrum. Indeed, in ordinary CFTs, given the scaling dimension and spin of an operator, two-point correlation functions are uniquely fixed by conformal invariance; however, as we already mentioned, the $O(n)$ model is a logarithmic CFT (logCFT). This implies that its correlation functions depend on logarithms of distances that, even at the level of two-point functions, can take one of many forms. As Cardy explained in \cite {Cardy:1999zp}, the $n=0$ theory is generically  logarithmic. He also studied the $O(n)$ model for other integer values of $n$ (and in any dimension) \cite{Cardy:2013rqg} and showed that certain operators also become logarithmic and their two-point functions do not have a simple power-law form. In 2d it was proven in \cite{Vasseur:2011fi} that the $O(n)$ model is logarithmic for a discrete infinite set of $n$'s -- those for which the central charge coincides with that of unitary minimal models. What we will find is that, in addition to logarithmic operators identified in the above papers, in two dimensions the model contains logarithmic operators for any value of $n$. In particular, we will see that even the conserved $O(n)$ current is a part of the logarithmic multiplet.

Since logarithmic operators are often thought as a result of tuning (more on this in section \ref{sec:log_limits}), their presence for a generic value of the model's only parameter $n$ may appear surprising. At the moment, we do not understand an underlying physical reason for appearance of logarithmic correlators; however, we would like to point out that similar situations have been recently observed in high energy physics. An example that appears closest in the spirit, is the worldsheet theory of strings propagating on a $\mathbb{Z}_N$ orbifold \cite{Dabholkar:1994ai}. As was shown in \cite{Witten:2018xfj}, analytic continuation of the annulus partition function in $N$ reveals the presence of logarithmic operators in this theory for generic $N$.\footnote{In the case of the $O(n)$ model calculation of the annulus partition function, which was done in \cite{Cardy:2006fg}, does not show logarithmic operators because all of them transform in non-trivial representations of $O(n)$ and have zero matrix elements with the boundary state picked in \cite{Cardy:2006fg}. Torus partition function, instead, contains these operators but does not exhibit any power-law features due to their logarithmic nature. } Similarly to what happens in the $O(n)$ model, when $N\to1$ the worldsheet CFT must become unitary, and consequently logarithmic operators should decouple. Another known family of logarithmic theories is provided by the Fishnet theories \cite{Gromov:2017cja}. It appears, however, that the nature of these theories is quite different because the Fishnet CFT is complex, while the $O(n)$ models that we study here are non-unitary but real theories (see \cite{Gorbenko:2018_1} for the distinction between the two). Finally, let us mention that 2d $Q$-states Potts model, which has many similarities in structure with the $O(n)$ model, also turns out to be a logarithmic theory for a generic value of $Q$. We will provide some details related to logarithmic properties of this model in section \ref{PottsLogs}. 

Part of our motivation for studying the $n\lesssim2$ critical $O(n)$ CFT is that for $n\gtr2$ this CFT becomes complex and, as explained in \cite{Gorbenko:2018_1,Gorbenko:2018_2}, controls the walking RG behavior of a unitary massive theory for integer $n\gtrsim2$. This type of RG flows is of interest for particle physics because they provide a natural way to generate a hierarchy of scales, and at the same time they control weakly-first-order phase transitions \cite{Wang:2017txt} in certain condensed matter systems. The $n>2$ CFT can be obtained from the better understood $n\leq2$ case by analytic continuation in $n$, and, consequently, the study of singularities present in the theory for $n\to2$, to which much of this paper is dedicated, is essential for performing this continuation. In this paper we will touch upon the $n\gtrsim2$ theory only briefly, deferring the detailed study to a future publication. Our preliminary investigation suggests a possible relation between these complex CFTs and periodic S-matrices recently discussed in the context of the S-matrix bootstrap program \cite{Cordova:2018uop,Cordova:2019lot}.

The outline of this paper is as follows. In section \ref{sec:recap} we briefly introduce the main features of logCFTs. Section \ref{sec:ZOn} summarizes the results of \cite{diFrancesco:1987qf} about the spectrum of the critical $O(n)$ model. These are used in section \ref{sec:crossing}, where we impose crossing symmetry in order to fix some of the OPE coefficients and four point functions of the theory, finding in the process the smoking gun that proves that the critical theory is logarithmic. Section \ref{sec:logs} is dedicated to structures of logarithmic multiplets; in particular, we have several non rigorous examples that helped us build intuition about them in subsection \ref{sec:log_limits}, which might be skipped by a reader well acquainted with the topic. The RG flow that connects the critical fixed point to the low-temperature fixed point is studied in \ref{sec:flow} in its perturbative regime, $n \lesssim 2$. We check in several cases that the conformal data for the critical fixed point, together with the rules of conformal perturbation theory, reproduce the correct spectrum for the low-temperature fixed point. In section \ref{sec:n->2} we show how logarithmic operators decouple and one recovers a unitary subsector of the theory when we take the limit $n \to 2$. Finally, we mention that also the two-dimensional critical Potts model is logarithmic in section \ref{PottsLogs} and we recap and address some open questions in \ref{sec:conclusions}. In particular, in section \ref{sec:Dangerous} we discuss the puzzles related to the singlet operator relevant in the low-T fixed point, and in \ref{sec:S-matrix} the $n>2$ regime.

 We summarize our main results here:
\begin{itemize}
	\item The critical $O(n)$ model and the low temperature fixed points are logarithmic CFTs for any value of $n$ in the range $-2 \le n \le 2$. This also applies to the two-dimensional critical Potts model at generic values of $Q$ for $0<Q<4$.
	\item The currents are part of a logarithmic staggered module. This implies that, at generic $n$, we don't have factorization into holomorphic and antiholomorphic currents and there is no enhancement of the symmetry, $O(n)\not \to O(n)_L\times O(n)_R$.
	\item When we take the $n \to 2$ limit, we recover a unitary subsector of the theory, whose correlation functions are reflection positive and have no logarithms. This is a result of highly non-trivial cancellations between operators.
\end{itemize}

\section{Logarithmic CFTs recap} \label{sec:recap}
Let us very briefly mention some properties of logarithmic CFTs that will be most relevant for our discussion. The field of logarithmic CFTs is significantly less developed than that of ordinary CFTs, however, it is still contains many more results than we can quote here. A very nice review is \cite{Hogervorst:2016itc} and several original papers that were very useful for us include \cite{Gurarie:1993xq,Gaberdiel:1996kx,Gaberdiel:1998ps,Kausch:2000fu}.

A logCFT is a CFT where the action of the dilatation operator $\calD$ cannot be diagonalized but can only be put into a Jordan block form. The simplest logarithmic multiplet of dimension $\Delta$ consists of two fields $A$ and $B$, of dimension $\Delta$, which under dilatations transform as
\beq
\calD \begin{pmatrix}
	A\\B
\end{pmatrix} = \begin{pmatrix}
\Delta & 1 \\0 &\Delta
\end{pmatrix}
\begin{pmatrix}
	A\\B
\end{pmatrix} \label{eq:DJordan}\,.
\eeq
In some appropriate normalization, the two fields have the following two point functions among themselves:
\beqg
\braket{A(x)A(0)}=\frac{\log \l(|x|^2 \mu^2\r)}{|x|^{2\Delta}}\,,\\
\braket{B(x)A(0)}=-\frac{\beta}{|x|^{2\Delta}}\,,\\
\braket{B(x)B(0)}=0\,. \label{eq:log2pf}
\eeqg
with $\beta$ a constant to be determined.
As it will be clear later, the operator $B$ forms an invariant submodule under the action of not only the dilation operator, but the whole conformal group (or Virasoro group in the case of two dimensions), so the multiplet formed by $A$ and $B$ is reducible. However, since we cannot write this as direct sum of invariant submodules, it is also indecomposable \cite{Rohsiepe:1996qj}.
Note also the appearance of the scale under the $\log$. This scale also cannot be removed, but it doesn't have any physical meaning because change in this scale can always be compensated by redefinition of $A\to A+ a B$, which leaves \eqref{eq:DJordan} invariant. The two point functions \eqref{eq:log2pf} are left invariant under such a transformation, and the same can be argued for higher point functions (section 2.6 of \cite{Hogervorst:2016itc}). Because of this, logarithmic fields are well-defined operators, unlike for example free massless bosons in 2d which also have logarithms in their two-point functions.

The structure of \eqref{eq:DJordan} can be generalized to the case where we have more than two fields that transform in the same Jordan block; the size of the Jordan block is called rank of the logarithmic multiplet. In the operators of the $O(n)$ model that we've considered, we've only encountered rank $2$ logarithmic multiplets, so we focus on this case. To guide the reader we will always denote by $B$ the operator which has a vanishing two-point function and by $A$ the operator with the purely logarithmic two-point function. 

The logarithmic multiplets we will encounter will not be the simplest possible ones; the basic structure \eqref{eq:DJordan} will be the same, but the action of special conformal transformation on $A$ and $B$ will be non trivial. The kind of logarithmic multiplets we will face are called \textit{staggered modules}, and a lot will be said about them in section \ref{sec:logs}. 
We will have to invoke some guesswork (compensated by multiple cross-checks) to determine their structure. Some amount of intuition about logarithmic multiplets can be gained from logarithmic free field theories, see for example \cite{Brust:2016gjy}, as well as taking some limits of ordinary CFTs \textit{\`a la} Cardy \cite{Cardy:2013rqg}.

A fundamental property of logCFTs is non-unitarity. This can be seen, for example, by computing the norm of states $A$ and $B$ in \eqref{eq:DJordan}, and see that the Gram matrix has a negative eigenvalue \cite{Hogervorst:2016itc}. One can also think of a non-diagonalizable dilatation operator as a non-diagonalizable Hamiltonian in radial quantization, which therefore cannot be hermitian. One of the main topics we will address in this work is how these negative norm states drop out of a sector of the theory when we need to recover a unitary theory for $n \to 2$, see section \ref{sec:n->2}. We expect this mechanism to be rather generic. In particular, it should operate in the critical model for $n\to1$.

\section{Operator content of the 2d $O(n)$ model}
\label{sec:ZOn}
Let us start with describing the spectrum of operators present in the critical and low-T $O(n)$ models. For this purpose we can use the partition function on the torus, which was calculated in \cite{diFrancesco:1987qf} with the use of the Coulomb gas formalism. This discussion is very similar to that in the Potts model, which we reviewed in some detail in \cite{Gorbenko:2018_2}, so we will be brief here. As usual, let us parametrize the torus by $q=e^{2 \pi i \tau}$ and $\bar q$. Then the spectrum can be read off from the expansion of the partion function in $q$ and $\bar q$:
\beq
Z_{O(n)}=(q \bar q)^{-\frac{c}{24}}\text{Tr}\,q^{L_0}\bar{q}^{\bar{L}_0}.
\eeq
To summarize the result of \cite{diFrancesco:1987qf} we need to introduce several quantities. First, we define the coupling $g$,\footnote{Since the theory is gaussian, the `coupling' is really related to the radius of the free boson.}
\beq
g=\arccos\l(\frac{n}{2}\r)\,,
\eeq
where $g\in[1,2]$ and $g\in[0,1]$ for critical and low-T theories correspondingly. In terms of $g$ the background charge $e_0$ is given by
\beq
e_0=g-1\,
\eeq
and the central charge of the $O(n)$ CFTs reads
\beq
\label{eq:c}
c=1-\frac{6e_0^2}{g}=1-6\frac{(1-g)^2}{g}\,.
\eeq
Finally, we will need the weights $x$, $\bar x$, which are labeled by electric and magnetic charges
\beq
x_{em},\bar{x}_{em}=\frac 14 (e/\sqrt{g}\pm m\sqrt{g})^2.
\eeq
In terms of these quantities, the expansion of the partition function reads \cite{diFrancesco:1987qf}
\beq
Z_{O(n)}=\frac{1}{\eta\bar{\eta}}\sum_{P\in\bZ}(q\bar{q})^{x_{e_0+2P,0}}+\frac{1}{\eta\bar{\eta}}\sum_{\substack{M,N=1\\N\text{ divides }M}}^\infty \Lambda(M,N) \sum_{\substack{P \in \bZ \\ P\wedge N=1}}q^{x_{2P/N,M/2}}\bar{q}^{\bar{x}_{2P/N,M/2}}. \label{eq:zhat}
\eeq
Coefficients $\Lambda(M,N)$ depend only on $n$ and their definition is given in Eq.~(3.24) of \cite{diFrancesco:1987qf}.\footnote{An alternative way of computing these coefficients is presented in \cite{Read:2001pz}.}

Note that nothing in \eqref{eq:zhat} suggests anything about the logarithmicity of the theory, but this should not be expected. Because of the trace, the torus partition function is not sensible to off diagonal terms in the dilatation operator. One could hope to see signs of logarithmicity by computing some twisted torus partition function or the cylinder partition function with non-singlet boundary states.

For now we ignore the subtleties associated with the presence of logarithmic operators, as well as of degenerate fields, which make distinction between primary and descendant operators somewhat complicated. In general, we may expect that terms $q^{x_{em}}{\bar q}^{{\bar x}_{em}}$ appearing in the expansion will correspond to primaries, which we will denote as $\calO_{e,m}$. Dimensions of those operators read
\beq
\label{hx}
h=x_{em}+\frac{c-1}{24},\qquad \bar h=\bar x_{em}+\frac{c-1}{24}\,.
\eeq
Here $e$ and $m$ are either rational numbers (for operators coming from the second term in $Z_{O(n)}$), or $e=e_0+2 P$, with $P \in \mathbb{Z}$ (the first term in $Z_{O(n)}$). 

\begin{figure}[!ht]
	\centering
	\includegraphics[width=0.46\textwidth]{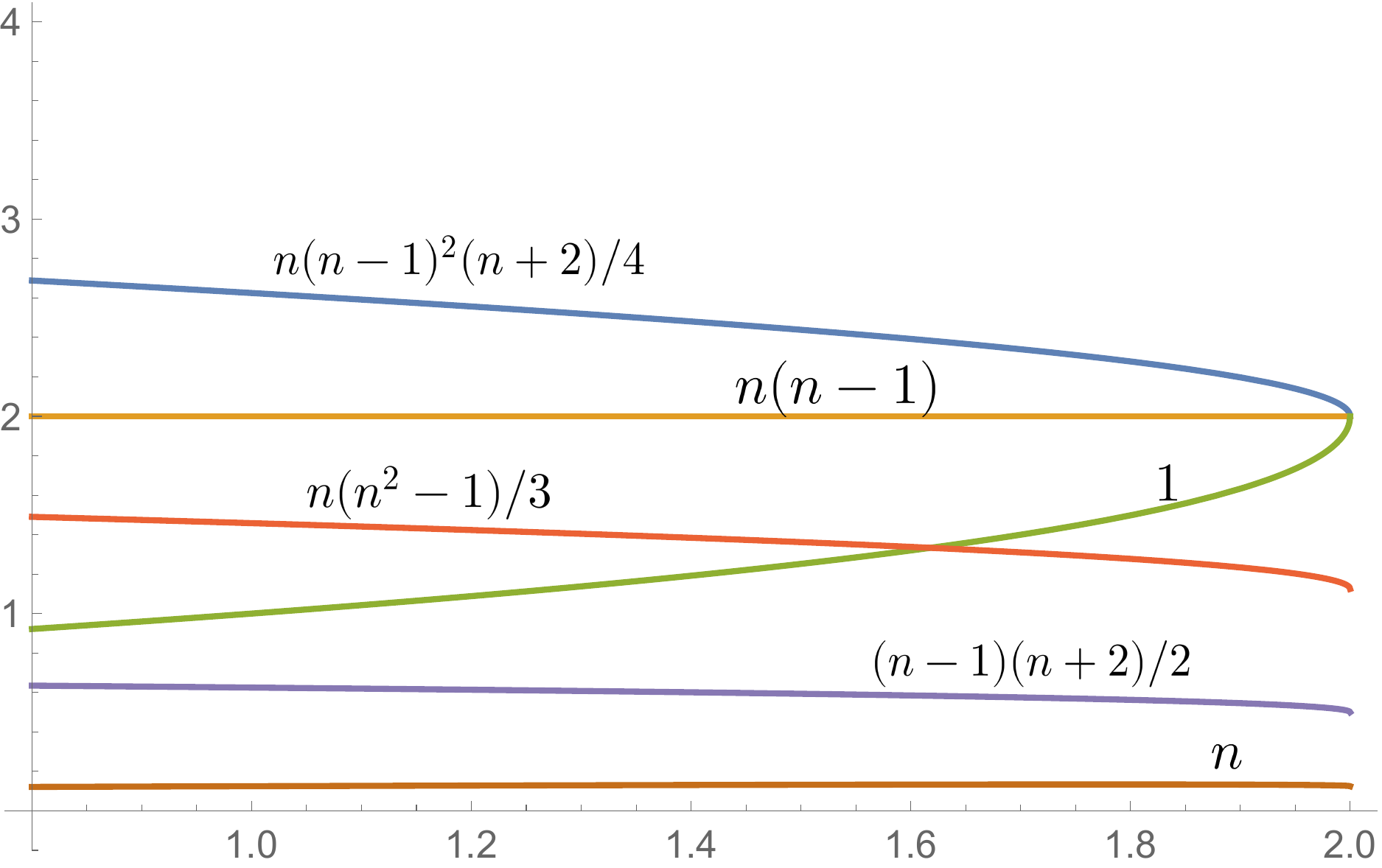}
	\hspace{0.5cm}
        \includegraphics[width=0.46\textwidth]{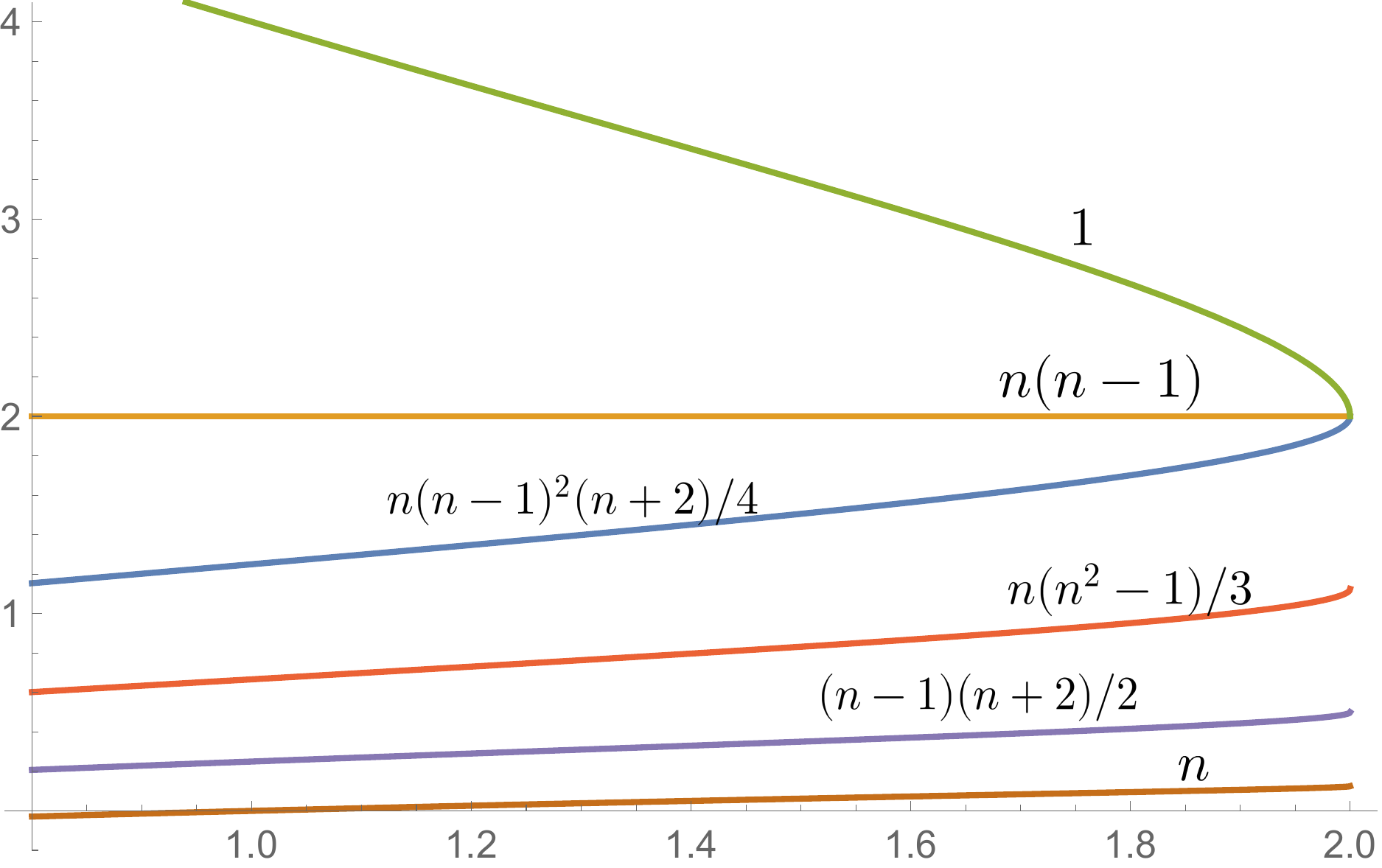}
	\caption{Dimensions and multiplicities of light scalar operators as functions of $n$ in the critical (left) and low-T fixed points (right).} 
		\label{fig:dim}
\end{figure}

For an illustration, we plot dimensions of the leading (for $n\lesssim2$) scalar operators as a function of $n$, and indicate their multiplicities, on Fig. \ref{fig:dim}. Non-trivial multiplicities indicate that at least some of the operators with corresponding dimensions transform non-trivially under $O(n)$. In fact, as conjectured in \cite{Gorbenko:2018_2} and proven in \cite{Binder:2019zqc}, $O(n)$ symmetry requires that all multiplicities correspond to dimensions of representations of $O(n)$, not necessarily irreducible.\footnote{It implies that coefficients with which dimensions of irreducible representations appear are positive $n$-independent integers.} Let us discuss this figure in some details. First, we identify the energy operator $\veps$, indicated in green, which is the only relevant singlet at the critical fixed point. It is irrelevant in the low-T theory and {it corresponds to $\calO_{em}$ with $e=e_0-2$ and $m=0$.} This operator will play a major role in our analysis. The most relevant operator in both models is the spin operator $\sigma$ ($e=0$, $m=1/2$), with multiplicity $n$, which corresponds to the vector representation of $O(n)$. Note that in both models there are $n(n-1)$ marginal scalar operators for any $n$. As we will see later, these is a pair of adjoint fields, related to the conserved $O(n)$ currents. These operators are not regular primaries.

Since $\veps$ is the only relevant singlet operator in the critical model, it should trigger the RG flow to the low-T fixed point. Under this flow a set of operators denoted in blue in figure \ref{fig:dim} transitions from being irrelevant to being relevant. It is thus important to determine whether any of these operators are singlets. This cannot be concluded just from the multiplicity, as the decomposition into irreducible representations is far from unique. However, by studying the $n\to2$ limit of the theory we will prove that exactly one of these operators is actually a singlet. This is the ``dangerously relevant'' operator, mentioned in the introduction and which properties we discuss in section~\ref{sec:Dangerous}. We note that equal number of relevant singlet operators in the IR and UV fixed points is unusual  from the Morse theory viewpoint on the RG flows \cite{Gukov:2015qea}.

Other operators will also play a role in our discussion. Note that all operator dimensions in two models merge at $n\to2$. In fact the operator dimensions in two theories go into each other upon analytic continuation around $n=2$.

\section{OPE coefficients from crossing} 
\label{sec:crossing}

Knowledge of the torus partition function gives us access to the spectrum of local operators in the theory, but tells us nothing about OPE coefficients. In this section, we will be using crossing symmetry of several four point functions to get some of the OPE coefficients. We will need these OPE coefficients in order to study the RG flow from the critical to the low-T fixed point to lowest order in $2-n$, as well as to study the $n \to 2$ recombination, and, in principle, the $n>2$ regime.

 Another reason for studying four point functions of the theory is that it gives more detailed information about the transformation properties of operators rather than the partition function.
Normally we are interested in the primaries of the theory, so what we should do is expand the partition function in characters, rather than just in $q$ and $\bq$. However, in a non-unitary theory operators can behave in a wilder way. For example, consider the marginal operators indicated in  orange in Fig.~\ref{fig:dim}.  In an unitary theory we would conclude that these operators are primaries, since they cannot be descendants of the currents or of the identity. By studying correlation functions of $J$ and $\veps$, we will see instead that half of these operators are descendants of the currents and have zero norm, while the other half are neither primaries nor descendants; together they form a logarithmic multiplet.

We will work to lowest order in conformal perturbation theory in an expansion in $2-n$. We will focus on the critical theory, although a very similar analysis can be performed for the low-T theory. It will be convenient to use the parameter 
\beq
\label{mn}
m(n)=\frac{\pi}{\cos ^{-1}\left(\frac{n}{2}\right)}\,,
\eeq
for which the limit $n \to 2$ corresponds to $m \to \infty$, $m\approx\frac{\pi}{\sqrt{2-n}}$. This parameter reminds of the counting parameter for minimal models: for example, the central charge of the critical theory is now given by $c=1-\frac{6}{m(m+1)}$, which is the same relation that holds for the $\mathcal{M}_{m,m+1}$ minimal models. While in unitary minimal models $m$ only takes integer values, here it takes continuous values in the range $(1,\infty)$. Unlike the unitary minimal models, $O(n)$ CFT is not rational, that is it has infinitely many primary fields. Nevertheless, some operators present in the $O(n)$ theory are also present in the minimal models and some properties carry over. For example, there are fields that are degenerate for any $n$, which includes, in particular, the energy operator $\veps$. In this regard the situation is somewhat similar to what happens in generalized minimal models \cite{Zamolodchikov:2005fy}, however, in our case the theory is non-diagonal because we have primaries with spin. As we already explained in the introduction, our theory has a microscopic formulation for any $n$, so the corresponding CFT is not just a formal analytic continuation from integer $m$'s.

The main feature we will use is that the energy operator $\veps$ is degenerate and has a null descendant. Following the seminal work of \cite{BPZ}, this means that correlators of $\veps$ satisfy a differential equation. Imposing this and crossing symmetry, we will be able to fix many of the OPE coefficients we're interested in. Indeed, using \eqref{hx} and \eqref{mn} we see that $h_\veps = h_{1,3}$, where we have defined the usual 

\beq
h_{r,s}=\frac{((m+1) r-m s)^2-1}{4 m (m+1)}\,. \label{eq:hKac}
\eeq
An operator whose dimension is $h_{r,s}$ for $r,s \in \mathbb{N}$ has a zero-norm descendant at level $rs$. However, since we are in a non-unitary theory, we need to make a distinction between zero-norm states and null states. By zero-norm operator we mean an operator which has zero two-point function with itself, while by null we mean an operator who, when inserted in any $n$-point function, always gives zero. In unitary theories, a zero-norm vector must also be null. The requirement of having strictly positive norms means that if we have a zero-norm operator, we need to mod it out of the theory; we can do this consistently only if any correlation function in which this operator is inserted is zero.
In a non-unitary theory, however, we can have zero-norm vectors which are not null; in our case, the simplest example is operator $B$ in \eqref{eq:log2pf}, for which $\braket{BB}=0$ but $\braket{AB}\neq 0$. Concerning the energy operator, therefore, having $h_\veps=h_{1,3}$ by itself only guarantees that a descendant is zero norm; we'd like to show that this descendant is actually null and therefore that the correlation functions of $\veps$ satisfy some useful differential equation.

We will use our only tool so far, the torus partition function, to settle this question. Our assumption is that, while zero-norm operators generically appear in the partition function, if an operator does not appear in the partition function for any $n$, it is not present in the theory. So we expand the partition function and count how many terms $\sim q^{h_{\veps}-c/24} \bq^{h_\veps+3-c/24}$ we have. If we were to find that the answer is two, we would conclude that one of the level three descendant of $\veps$ is null. The situation is just a bit more complicated, but the conclusion remains the same.

We find in fact that there are $3+\Lambda(2,2)=2+\frac{n(n-1)}{2}$ operators of dimension $(h_\veps,h_\veps+3)$. This is the only decomposition of this multiplicity in terms of irreps of $O(n)$ with non-negative and $n$ independent coefficients, property which is required by the categorical $O(n)$ symmetry \cite{Binder:2019zqc}.  Since we only have two operators with weight $(h_\veps,h_\veps+2)$, which is the right number of level-two primaries of $\veps$,
we conclude that, at $(h_\veps,h_\veps+3)$, we have $\frac{n(n-1)}{2}$ primaries, transforming in the antisymmetric of $O(n)$, and two descendants of $\veps$. One of its level three descendant does not appear, and we conclude that it's null.

One can check explicitly that this null descendant is
\beq
\left( L_{-3}-\frac{2 (m+1)}{3 m+1}L_{-1}L_{-2}+\frac{(m+1)^2}{2 m (3 m+1)} L_{-1}^3\right) \veps =0 \label{eq:nullOp}
\eeq
This means that correlators involving an insertion of the operator $\veps$ satisfy the differential equation \cite{BPZ}
\beq
\left(\calL_{-3} -\frac{2 (m+1)}{3 m+1}\calL_{-1}\calL_{-2}+\frac{(m+1)^2}{2 m (3 m+1)} \calL_{-1}^3\right) \braket{\veps(z) \phi_1(z_1) \ldots \phi_n{(z_n)}}=0 \label{eq:BPZ4pf}
\eeq
where
\beq
\calL_{-k}=\sum_{i=1}^n \left(\frac{(k-1)h_i}{(z_i-z)^k}-\frac{1}{(z_i-z)^{k-1} }\pd_{z_i} \right) \,. \label{eq:Ldiff}
\eeq

Another reason for studying correlation functions of the operators $\veps$ is that, as already mentioned, under perturbations by $\veps$, the critical point can flow to a IR low-temperature fixed point. The flow is perturbative for $n \lesssim 2$, and in section \ref{sec:flow} we will check that the conformal data we have for the two fixed points agrees with conformal perturbation theory. In order to do so, we need OPE coefficients of operators with $\veps$ at $n=2$. We refer to section  \ref{sec:flow} for the details.

\subsection{The energy operator 4pt function}

Let's start by considering the correlator $\braket{\veps \veps \veps \veps}$, and, for the moment, let's focus on the holomorphic dependence only.
\beq
\braket{\veps(z_1) \veps(z_2) \veps(z_3) \veps(z_4)}=\frac{1}{(z_{12} z_{34})^{2h_\veps}}f(z)\,,
\eeq
where $z_{ij}=z_i-z_j$, and $z$ is the usual cross ratio
\beq
z=\frac{z_{12}z_{34}}{z_{13}z_{24}}\,.
\eeq
Acting with the differential operator \eqref{eq:nullOp} on one of the operator insertions, we obtain the following third order differential equation for $f(z)$
\beq
f'''(z)+c_2(z) f''(z)+c_1(z) f'(z)+c_0(z) f(z)=0 \label{eq:diffeq4eps}
\eeq
where the coefficients are explicitly given by
\beqa
c_0(z)&=\frac{4 (m-1)^2 m (z-2)}{(m+1)^3 (z-1)^3 z}\\
c_1(z)&=\frac{2 \left(m^2+m-6\right) z+2 (m (3-2 m)+3) z^2-6 m+6}{(m+1)^2 (z-1)^2 z^2}\\
c_2(z)&=\frac{2 ((m+3) z+m-3)}{(m+1) (z-1) z}   \label{eq:diffeq4eps_coeffs}
\eeqa

We are after the Virasoro blocks of the operators exchanged in this four point function, which are solutions to the differential equation \eqref{eq:diffeq4eps}.\footnote{{Here we are using high energy terminology, which might be unfamiliar to more statistically mechanics oriented readers: when considering a four point function $\braket{\phi_1(0) \phi_2(z,\bz) \phi_3(1) \phi_4(\infty)}$ by exchanged operators in the s-channel we mean the operators that appear in both the $\phi_1 \phi_2$ and $\phi_3 \phi_4$ OPE. Exchanged operators in the t-channel are instead those appearing in both the $\phi_1 \phi_4$ and $\phi_2 \phi_3$ OPE.}}
For generic values of $m$, we cannot find solutions to \eqref{eq:diffeq4eps} in closed form. It will be enough to find the Virasoro blocks approximately as a series expansion in $z$, by keeping a finite number of terms. We will see later that in the $m \to \infty$ limit ($n \to 2$) we are able to solve the differential equation exactly.

The Virasoro block corresponding to the exchange of an operator $\phi$ with holomorphic weight $h_\phi$ behaves as
 \beq
\calV_{\phi}(z)=z^{h_\phi} \sum_{n=0}^\infty c_n z^n\,, \quad c_0=1\,. \label{eq:VBlock}
\eeq
We look for solutions to \eqref{eq:diffeq4eps} who behave as $z^{h_\phi}$ for small $z$. Solving the indicial equation,
\begin{equation}
	h_\phi^3 (m+1)^2+h_\phi^2 \left(-5 m^2-2 m+3\right)+h_\phi \left(4 m^2-6 m+2\right)=0
\end{equation}
we see that the allowed values of $h_\phi$ are $0$, $h_\veps$ and $h_{\veps'}\equiv h_{1,5}$. This is to be expected, as the fusion of two operators with dimension $h_{r,s}$, with $r,s$ positive integers, is known \cite{BPZ}. For generic values of $m$, these three numbers do not differ by an integer, and it's straightforward to identify which solution of the differential equation is the Virasoro block for each operator. 
 We will see later on that when two roots of the indicial polynomial differ by an integer, identifying the correct Virasoro block requires more work, and in some case we can obtain solutions which are not series in $z$, but depend on $\log z $ as well.
 
Since we are looking at a four point function of scalars, the Virasoro blocks are the same in the holomorphic and in the antiholomorphic channel. Therefore the four point function in the s-channel will look like
\beq
\braket{\veps(0)\veps(z,\bz)\veps(1)\veps(\infty)}=\frac{1}{(z \bz)^{2h_\veps}}\left(\calV_\id(z) \calV_\id(\bz) +\lambda_{\veps \veps \veps}^2 \calV_\veps(z)  \calV_\veps(\bz)+\lambda_{\veps \veps \veps'}^2 \calV_{\veps'}(z)  \calV_{\veps'}(\bz)\right) 
\eeq
The primary operators we can exchange are the identity operator, the energy operator itself, with weight $(h_\veps,h_\veps)$, and another operator which we call $\veps'$, with dimension $(h_{\veps'},h_{\veps'})$. They all have spin zero. Notice that, since for generic values of $m$ all the weights of the exchanged operators differ by numbers which are not integers, we cannot exchange primaries with spin.

Since we are looking at the correlator of four identical operators, the t-channel expansion of the four point function looks the same with $z \to 1-z$ and $\bz \to 1-\bz$. At this point we impose crossing symmetry in order to fix the OPE coefficients. In practice, for a given value of $m$ we compute the Virasoro blocks easily to the first few hundred orders in $z$ and then impose crossing symmetry by requiring that derivatives with respect to $z$ and $\bz$ at the crossing symmetric point $z=\bz=1/2$ are the same in the s-channel and t-channel expansion. This uniquely fixes the OPE coefficients to the usual diagonal minimal model solution, originally obtained by \cite{Dotsenko:1984ad} (see for example appendix A of \cite{Poghossian:2013fda} for more compact expressions). This is to be expected: our theory is not a minimal model, but the operator $\veps$ is part of the Kac table, and all operators exchanged in this correlation functions are scalars, so crossing symmetry fixes the OPE coefficients to be the same of a diagonal minimal model. We will see other correlation functions where we can exchange spinning operators\footnote{By spinning operators we mean operators with nonzero spin, $h \neq \bar h$.} and where we cannot use directly results from the diagonal minimal models. Nevertheless, most of the OPE coefficients we found are related to the minimal model ones in some way, see Appendix~\ref{app:minimalOPE} for these relations.

We can also work in the $m\to \infty$ limit, where the coefficients \eqref{eq:diffeq4eps_coeffs} take a simpler form. We can solve equation \eqref{eq:diffeq4eps} exactly, but now the roots of the indicial equation differ by integers (they are 0, $h_\veps=1 $ and $h_{\veps'}=4$). Let's say for example that we are interested in the Virasoro block of the identity. Requiring this to be have the form \eqref{eq:VBlock} does not fix the block uniquely. We have two undetermined constants, which are related to the fact that we can always add $\calV_{\veps}$ and $\calV_{\veps'}$ to $\calV_\id$ and it will still satisfy \eqref{eq:VBlock}. The way to fix this is to work out the first coefficients (up to $z^4$) of the identity block using the explicit action of the generators $L_{-n}$, see for example formula (6.190) and (6.191) in \cite{DiFrancesco_book}. Practically, this is quite complicated to do, so it's simpler to compute the first four coefficient of the identity block for generic values of $m$, where $h_\veps$ and $h_{\veps'}$ are not integers and no ambiguity is present, and then take the limit $m \to \infty$. There is only one solution to the differential equations in the $m \to \infty$ limit which has this specific small $z$ behavior, and this is the identity Virasoro block.

Explicitly, at $n=2$, we have
\beqg
\calV_\id(z)=\frac{3-6z+9z^2-6z^3+z^4}{3 (1-z)^2}\\
\calV_\veps(z)=\frac{z \left(4-6 z+4 z^2-z^3\right)}{4 (1-z)^2}\\
\calV_{\veps'}(z)=\frac{z^4}{(1-z)^2}\\
\eeqg
and the OPE coefficients turn out to be
\beq
\lambda_{\veps \veps \veps}=\pm \frac{4}{\sqrt{3}} \qquad \lambda_{\veps \veps \veps'}=\pm \frac{\sqrt{5}}{3}
\eeq
which agrees with the $m\to \infty$ limit of the diagonal minimal model OPE coefficients. The plus or minus ambiguity comes from the fact that this specific four point function is dependent only on the squares of the OPE coefficients. We choose the plus signs, and we would recover the minus signs by just redefining $\veps \to -\veps$ and $\veps' \to -\veps'$.

Besides working at $m=\infty$, we can also obtain a closed form expression of a few subleading corrections for the correlation functions in a $\frac{1}{m}$ expansion. In the case of $\braket{\veps \veps \veps \veps}$, we did this to $O\left( \frac{1}{m^2}\right)$, and checked agreement with numerical results for large but finite $m$.

\subsection{The spin operator}
One of the most important operators of our theory is the spin operator $\sigma$, and we are after the $\lambda_{\sigma \sigma \veps}$ OPE coefficient. For generic values of $n$, $\sigma$ is not a degenerate operator,\footnote{We have that $h_\sigma=h_{\frac{m-1}{2},\frac{m+1}{2}}$, so $\sigma$ is degenerate only for odd values of $m$.} but we will look at the four point function $\braket{\sigma \sigma \veps \veps}$ and use again the fact that $\veps$ has a null descendant at level three for all values of $n$.
The situation here is a bit more complicated than before because, when we look at $\braket{\sigma \sigma \veps \veps}$, the s and t-channel expansion correspond to two different OPEs, but the logic is the same as before.

One complication is that $\sigma$ transforms non-trivially under $O(n)$. For integer $n$, this is not a big deal since we can just put corresponding $O(n)$ representation indices and contract them in a necessary way in correlation functions, {\it i.e.} $\langle \sigma^a \sigma^b\rangle \sim \delta^{ab}$. For continuous $n$ such indices do not have a clear meaning. Using category theory, \cite{Binder:2019zqc} developed a machinery which allows to deal with operators transforming non-trivially under continuous-$n$ $O(n)$ symmetry in a rigorous way. In particular, they generalized the notion of operators transforming in irreducible representations. For most of this work we will deal with correlation functions that have two operators transforming in some irreducible representation of $O(n)$ and singlet correlators. Thus, independently of the channel in which we look at the correlator, its categorical symmetry structure is the same and formally factors out in all our equations. Based on this, in what follows we will simply not write any indices on the operators. Naturally, we will keep trace of the fact that OPE of a singlet operator with an operator in some irreducible representation can only produce operators in this representation. Needless to say, a more detailed study of the continuous $n$ theory would require more usage of the machinery of \cite{Binder:2019zqc}. 

Let's work first at some generic value of $m$. In the s-channel, looking again at the holomorphic part only,
\begin{equation}
\braket{\sigma(z_1) \sigma(z_2) \veps(z_3) \veps(z_4)}=\frac{1}{z_{12}^{2h_\sigma}z_{34}^{2h_\veps}} f_s(z)\,.
\end{equation}
By acting with the differential operator \eqref{eq:nullOp} on, say, the energy operator at $z_3$, we obtain a differential equation for $f_s(z)$. The roots of its indicial equations are again $0$, $h_\veps$ and $h_{\veps'}$. This was to be expected, since in the s-channel expansion we are doing the OPE between $\veps$ and $\veps$, and we know this OPE from the previous section,
\beq
\veps \cdot \veps = \id +\veps +\veps'\,.
\eeq
Since the operator $\sigma$ is not degenerate, the OPE $\sigma \cdot \sigma$ can contain, in principle, an infinite number of primaries, but because of the simplicity of the $\veps \cdot \veps$ OPE, we only exchange three operators in the s-channel.

In the t-channel expansion, we have
\begin{equation}
\braket{\sigma(z_1) \sigma(z_2) \veps(z_3) \veps(z_4)}=\frac{1}{z_{23}^{h_\sigma+h_\veps}z_{14}^{h_\sigma + h_\veps}} \left(\frac{z_{34	}}{z_{31}} \right)^{h_\sigma-h_\veps} \left(\frac{z_{24	}}{z_{34}} \right)^{h_\veps-h_\sigma} f_t(1-z)\,.
\end{equation}
and we get a different differential equation for $f_t(z)$. The roots of the indicial equation are $h_\sigma$, $h_{\sigma'}=h_{\frac{m-1}{2},\frac{m+1}{2}-2}$ and $h_{\sigma''}=h_{\frac{m-1}{2},\frac{m+1}{2}+2}$. It's crucial to notice that $h_{\sigma''}-h_{\sigma'}=1$ for all values of $m$, making the process of identifying the correct Virasoro holomorphic blocks not as straightforward as before. Requiring the $\calV_{\sigma}$ Virasoro block to be of the form \eqref{eq:VBlock} fixes uniquely the corresponding $c_n$'s; the same works for $\sigma''$. Getting the right $\calV_{\sigma'}$ is more complicated, since again we can always shift $\calV_{\sigma'}$ by a term proportional to $\calV_{\sigma''}$ while preserving \eqref{eq:VBlock}.

At this point, we are forced to look explicitly at the coefficient of the first subleading term of the $\calV_{\sigma'}$. After doing the computation,\footnote{We are lucky here because $h_{\sigma''}-h_{\sigma'}=1$. In general, if $h_{\sigma''}-h_{\sigma'}=k$, we would have to look at the $k$-th subleading term in the small $z$ expansion of the conformal block. The explicit computation of these terms becomes complicated very quickly; for the case at hand formula (6.190) of \cite{DiFrancesco_book} is enough, but in general we found the function \texttt{BlockCoefficient} of the Mathematica package \cite{VirasoroPackage} useful. Notice that, in formula (6.190) of \cite{DiFrancesco_book} we need to put $h_1=h_4=h_\veps$ and $h_2=h_3=h_\veps$ because we are looking at the t-channel expansion while they are doing a s-channel one.} we find $\calV_{\sigma'}(z)$ as the unique solution of the differential equation that, for small $z$, behaves as
\beq
\calV_{\sigma'}(z)=z^{h_\sigma'} \left[ 1+\frac{(h_{\sigma'}+h_{\sigma}-h_{\veps})^2}{2h_{\sigma'}}z+\ldots \right] 
\eeq

While in the s-channel we cannot exchange spinning primaries, the same isn't true in the t-channel. We have to combine the holomorphic and the antiholomorphic blocks so that we exchange only operators which are present in the theory. From the partition function we can check that the theory contains, besides the operator $\sigma$ itself, a primary operator $V$ with dimension $\Delta=h_{\sigma'}+h_{\sigma''}=2h_{\sigma'} +1$ and spin $1$, but no spinless primaries with dimension $\Delta=2h_{\sigma'} $ or $\Delta=2h_{\sigma''} $. We also notice that the multiplicity of $V$ is $n$, and is consistent with $V$ being a vector of $O(n)$ and with $V$ being exchanged in the $\veps \cdot \sigma$ OPE. Now we impose crossing symmetry
\beqg
\braket{\sigma(0)\sigma(z,\bz)\veps(1)\veps(\infty)}=\frac{1}{(z \bz)^{2 h_\sigma}} \left(\calV_\id(z) \calV_\id(\bz) +\lambda_{\veps \veps \veps}\lambda_{\sigma \sigma \veps} \calV_\veps(z)  \calV_\veps(\bz)+\lambda_{\veps \veps \veps'}\lambda_{\sigma \sigma \veps'} \calV_{\veps'}(z)  \calV_{\veps'}(\bz)\right) =\\
=\frac{1}{((1-z) (1-\bz))^{h_\veps+ h_\sigma}} \left[\lambda_{\sigma \sigma \veps}^2 \calV_{\sigma}(1-z) \calV_{\sigma} (1-\bz)  + \lambda_{\sigma \veps V}^2 \left( \calV_{\sigma'}(1-z) \calV_{\sigma''} (1-\bz) +\calV_{\sigma''}(1-z) \calV_{\sigma'} (1-\bz) \right) \right]\,.
\eeqg
We should not expect crossing to fix the OPE coefficients to be the same as the diagonal minimal models, since we are exchanging fields with spin, and indeed it is not the case.\footnote{We compute the OPE coefficients at several values of $m$ and find that $\lambda_{\sigma \sigma \veps}$ has a relative minus sign compared to what one expects in the diagonal minimal models, see appendix \ref{app:minimalOPE}.}
We can again do the computation in the $m\to \infty$ limit, and find that
\beqg
\lambda_{\sigma \sigma \veps}=-\frac{1}{8\sqrt{3}}+\ldots \qquad \lambda_{\sigma \sigma \veps'} =\frac{1}{24576 \sqrt{5}}+\ldots \qquad \lambda_{\veps \sigma V} = \pm \sqrt{\frac{1}{3}}+\ldots
\eeqg
The ambiguity in the last OPE coefficient follows again from the possibility of redefining $V \to -V$. 
The explicit formula of the $n=2$ correlator is in appendix \ref{app:n2explicit}.

\subsection{The currents} \label{sec:cross_currents}

Another important low-lying operator is the current $J_\mu$. It is in the adjoint representation of $O(n)$, as can be seen from its multiplicity $n(n-1)/2$, but as indicated above we drop corresponding indices. Since we will be looking at the four-point functions with two currents and two singlets there is a unique $O(n)$ tensor structure one can write for any $n$, and hence a single function of coordinates which can be associated to this correlator. As always in 2d, it is convenient to consider separately the components $J$ and $\bar J$; however, unlike in the usual unitary theories, these components will not be holomorphic or anti-holomorphic. Instead, we will find a single linear combination which is conserved. We will start by considering first the $\braket{J J \veps \veps}$ four point function. We repeat the by now usual procedure. In the s-channel we again exchange the identity operator, $\veps$ and $\veps'$. The only exception is that, since the operator $J$ has spin, the holomorphic and antiholomorphic blocks are different and satisfy two separate differential equation. The t-channel expansion, however, is more complicated.

Let's start by taking the holomorphic differential equation for the t-channel. The roots of the corresponding indicial equation are $0$, $1$ and $h_\veps+3$. However, we can only find two independent solutions as power expansions in $z$, which are $\calV_1$ and $\calV_{h_\veps+3}$. The novelty is that the differential equation admits a logarithmic solution as well, of the form
\beq
\widetilde \calV_0=1+\sum_{n=1}^\infty (c_n+d_n \log z)z^n\,. \label{eq:logBlock}
\eeq
Logarithms of the cross ratio appearing in a Virasoro (or conformal) block are related to the exchange of a logarithmic operator \cite{Gurarie:1993xq,Hogervorst:2016itc}.
The same happens for the antiholomorphic differential equation, whose indicial equation has roots $0$, $1$ and $h_\veps$. The corresponding solution is also of the form \eqref{eq:logBlock}. We have again the ambiguity of identifying what is the correct Virasoro block $\calV_0$, since we can always shift it by $\calV_1$, but now that we have logarithms around, the procedure of finding the correct Virasoro block is more complicated. The main reason is that, as we will explain in section \ref{sec:logs}, we can always shift one operator in the logarithmic multiplet by the other, and different shift will correspond to different definition of conformal blocks. This is why, for now, we denote a generic logarithmic solution of the differential equation as $\widetilde \calV_0$.

The expansion of the correlator in the t-channel looks generically\footnote{It's good to remember that we are dealing with a non-unitary theory. For example, the OPE coefficient $\lambda_{J \bJ \veps}$ would have to be zero in a unitary theory, unless $h_\veps=1$; this follows from imposing $\pd \bJ=\bpd J=0$. However, in our case the theory is non unitary and $\pd \bJ\neq0$, $\bpd J\neq0$, and $\lambda_{J \bJ \veps}$ turns out to be non vanishing.}
\beqg
\braket{J(0)J(z,\bz)\veps(1)\veps(\infty)}=\frac{-1}{(1-z)^{h_\veps+1}(1-\bz)^{h_\veps}}\left( \lambda_{J J \veps}^2 \calV_1 \widetilde{\bcalV}_{0}  + \lambda_{J \bJ \veps}^2 \widetilde{\calV}_{0} \bcalV_1 + \widetilde{\lambda}_{J \veps M}^2 \calV_1 \bcalV_1 +  \lambda_{J \veps \calO_{3,1}}^2 {\calV}_{h_\veps+3} \bcalV_{h_\veps}\right)  \label{eq:JJtchan}
\eeqg
By $M$ we mean some linear combination of the marginal operators $A$ and $B$, that will be discussed in section \ref{sec:currents}. 
The meaning of the ``tilde'' on $\widetilde{\lambda}_{J \veps M}$ is the same as above: this quantity changes upon choosing a different solution $\widetilde{\calV}_0$. This ambiguity will be fixed in section \ref{sec:currents}, while for now it suffices to use the schematic expression \eqref{eq:JJtchan}, which contains the information about the dimensions of the operators exchanged.

In \eqref{eq:JJtchan} we did not include a term $\calV_0 \bcalV_0$, even though this would naively look fine, given that it has integer spin. However, this would have terms like like $\log (1-z) \log (1-\bz)$, which are not single valued as we send $z$ and $\bz$ around $1$ in opposite directions. We also remark that in order to have single valuedness, we should be able to write all log terms as $\log (1-z)(1-\bz)$. This will not happen for generic values of $\lambda_{J J \veps}$ and $\lambda_{J \bJ \veps}$, but we see that crossing naturally fixes these OPE coefficients to values which make the four point function single valued.

Once we made an arbitrary choice for $\widetilde{\calV}_0$ and $\widetilde{\bcalV}_0$, crossing fixes all our OPE coefficients to a unique solution. Different choices of $\widetilde{\calV}_0$ and $\widetilde{\bcalV}_0$ correspond to different values of $\widetilde{\lambda}_{J \veps M}$, but the resulting crossing symmetric four point function is invariant under this ambiguity. 

We don't expect to see logarithms at $n=2$, and indeed one can check that, in the $n \to 2$ limit, the coefficients $d_n$ go to zero. At $n=2$ we can solve the differential equations exactly, and we can solve our ambiguity in identifying the correct $\calV_0$, since we don't have logs around anymore. We find\footnote{Throughout this work we normalize the currents to have a two point function $\braket{J J}=-z^{-2}$ for every $n$. In the $O(2)$ model, the current is $\pd \varphi$ and it indeed has a two point function $\sim -z^{-2}$. Another natural normalization of the currents, which makes contact with the average area of loops, is given in \cite{cite-key}.} %
\begin{equation}
	\langle J(0)J(z)\veps(1)\veps(\infty) \rangle=-\frac{1}{z^2}\frac{3-6z+9z^2-6 z^3+z^4}{3(1-z)^2}+O\left(\frac{1}{m}\right) \label{eq:JJn2}
\end{equation}
This allows us to read off the $n=2$ OPE coefficients $\lambda_{JJ \veps}$ and $\lambda_{J \bJ \veps}$, among other. It can be seen that, at this order, $\lambda_{JJ \veps}$ vanishes; since we're interested in the first non-vanishing order, we can compute the $O\left(\frac{1}{m}\right)$ correction to \eqref{eq:JJn2} explictily. We find
\beq
 \label{eq:JJe_OPE}
	\lambda_{J J \veps}=\frac{2}{\sqrt{3}m}+O\left(\frac{1}{m^2}\right) \qquad \lambda_{J \bJ\veps}=\pm \frac{1}{\sqrt{3}}+O\left(\frac{1}{m}\right)
\eeq

Something else worth remarking  about this correlation function is that, while $\braket{J \bpd J \veps \veps}=0$ at $n=2$, this is not true for generic $n$, meaning that $\bpd J \ne 0$. For example, in the $z, \bz \to 0$ limit we have
\beq
\langle J(0)J(z)\veps(1)\veps(\infty) \rangle=\frac{1}{z^2} \left( -1+ \lambda_{J J \veps} \lambda_{\veps \veps \veps} (z \bz)^{h_\veps}+\ldots \right)  \label{eq:JJinfinity}
\eeq
and we've seen that both $\lambda_{\veps \veps \veps}\ne0$ and $\lambda_{J J \veps}\ne0$ for generic values of $n$.
Only in the $n \to 2$ limit $\lambda_{J J \veps}$ vanishes and all the $\bz$ dependence drops out of $\braket{JJ\veps\veps}$.
This implies that, for generic $n$, the holomorphic and the antiholomorphic currents are not separately conserved. One linear combination of $\bpd J$ and $\pd \bJ$ will be zero, since there is a conserved current $ J^\mu$, but the orthogonal combination will be a non-trivial operator, which has zero norm but is not null. We will discuss this in detail in the next section.\footnote{We note that non-holomorphicity of currents in closely related 2d logCFTs has been previously observed in \cite{Read:2001pz}. Adjoint marginal scalars are also present in these theories.} 

One can also study the correlator $\braket{J \bJ \veps \veps}$. The only new feature of this correlator is that nowhere we exchange the identity. In previous correlators, we would fix the overall normalization by setting to one the coefficient in front of the identity Virasoro block. Here we cannot do that, and therefore we will not get OPE coefficients from crossing, but only ratios among them. In particular, it could always be that the four point function is zero. We can check explicitly, however, that in the $O(2)$ model this correlator is non zero. The ratio of OPE coefficients we find from this correlation function is compatible with the values of $\lambda_{\veps \veps\veps}$, $\lambda_{JJ\veps}$ and $\lambda_{J \bJ \veps}$ found earlier, and provides an independent consistency check.

\subsection{$\calO_{0,k/2}$ operators}
\label{sec:O0k}
Our theory contains a series of scalar operators $\calO_{0,k/2}$ for $k \in \mathbb{N}$, with dimensions $2h_{k\frac{m-1}{2},k\frac{m+1}{2}}$.
These operators are called $k$-leg operators or watermelon operators, and have a clear geometric interpretation in the loop formulation of the model as of an operator which originates $k$ lines, see for example \cite{PhysRevLett.58.2325} or \cite{Jacobsen2012}.
For $k=1$, this is the spin operator we already discussed. We study the correlator $\braket{\calO_{0,k/2}\calO_{0,k/2}\veps\veps}$, and we find in general that, for even $k$, the OPE $\calO_{0,k/2} \cdot \veps$ contains logarithmic operators, while for odd $k$ it doesn't (we checked this explicitely up to $k=6$). For even $k$, the logarithms appear at level $k$ in some Virasoro block, i.e. the BPZ differential equation has a solution of the form
\beq
\calV_{h_l}(z)=z^{h_l}\left( \sum_{n=0}^\infty c_n z^n+\sum_{n=k}^\infty d_n  z^n\log z\right) 
\eeq
and 
\beq
h_l=h_{k\frac{m-1}{2},k\frac{m+1}{2}-2}=h_{k/2,2}
\eeq
Notice that, based on its scaling dimension, we might expect this operator to have a null descendant at level $k$. As it will be explained in the next section, this operator is not null but zero norm, and this is closely related to the fact that logarithms appear at level $k$ of the Virasoro block. This happened as well for the $\braket{JJ\veps\veps}$, where we naively could expect $\bpd J$ to be null and logs appear at level 1 in the Virasoro block of $J$.

Concerning the OPE coefficients, we find that $\lambda_{\calO_{0,k/2} \calO_{0,k/2} \veps}=-\frac{k^2}{8\sqrt{3}}$ (checked explicitly up to $k=6$).

\section{Some details about logarithmic operators} \label{sec:logs}

\subsection{Logarithmic CFTs as limits} \label{sec:log_limits}
A standard way to get a logCFT is to take a family of ordinary CFTs, dependent on some parameter $\gamma$, such that for a given $\gamma=\gamma_0$ some operators have the same scaling dimension. Then, by carefully defining some observables, we find a logarithmic theory in the $\gamma \to \gamma_0$ limit. This was shown to happen generically in any dimension when one takes, for example, a $S_Q$ or $O(n)$ symmetric theory, with $Q$ or $n$ playing the role of $\gamma$, and then sends it to some integer value \cite{Cardy:2013rqg,Vasseur:2013baa}.

This is not what we expect to happen for the logarithmic operators we identified above, since we have only one parameter, $n$, and the theories are logarithmic for all values of $n$. However, we can artificially deform the $O(n)$ theory, by adding some extra parameter $\gamma$, such that for finite $\gamma$ our theory is not logarithmic, and for $\gamma \to 0$ we recover the $O(n)$ model and we see how logarithms arise. Clearly we should not rely too much on this procedure, as in principle we don't know if a consistent CFT exists at finite values $\gamma$, but we will use this approach to build intuition on what the logarithmic structure of our operators looks like. Eventually, we will forget about the trick involving an extra parameter and obtain the results more rigorously in section \ref{sec:currents}.

\subsubsection{Ordinary modules}
For now we work in a general number of dimensions and discuss the simplest example of logCFT we know of, taken from \cite{Hogervorst:2016itc}.
Assume that we have an ordinary, non-logarithmic, CFT with two primary fields, $\chi_1$ and $\chi_2$, with dimension $\Delta+\gamma$ and $\Delta$ respectively. We also take their two point function to be
\beq
\braket{\chi_1(x)\chi_1(0)}=\frac{-1}{|x|^{2(\Delta+\gamma)}}\qquad \braket{\chi_2(x)\chi_2(0)}=\frac{1}{|x|^{2\Delta}}
\eeq
Clearly, nothing weird happens for $\gamma \to 0$, at least from the point of view of two point functions of these operators, but we can build an operator that will be logarithmic in the $\gamma \to 0$ limit. Let's form the combination
\beqg
A_\chi=\frac{1}{\sqrt{\gamma}}(\chi_1+\chi_2) \qquad B_\chi=-\sqrt{\gamma} \chi_2\,. \label{eq:ABordinary}
\eeqg
For non-zero $\gamma$, $A_\chi$ is not an operator of definite scaling dimension, but we are interested in the $\gamma \to 0$ limit, where the two point functions look like
\beqg
\braket{A_\chi(x)A_\chi(0)}=\frac{\log |x|^2}{|x|^{2\Delta}}\\
\braket{B_\chi(x)A_\chi(0)}=-\frac{1}{|x|^{2\Delta}}\\
\braket{B_\chi(x)B_\chi(0)}=0\,.
\eeqg
The dilatation operator $D$ acts non diagonally on $A$
\beqa
D A_\chi &= \frac{1}{\sqrt{\gamma}}(\Delta+\gamma)(\chi_1+\chi_2) -\sqrt{\gamma} \chi_2 \xrightarrow[\gamma \to 0]{}\Delta A_\chi+B_\chi\\
D B_\chi &= \Delta B_\chi
\eeqa
This shows that it's possible to have logarithmic operators for $\gamma \to 0$.\footnote{The reader might wonder why we went through the effort of defining operators $A_\chi$ and $B_\chi$, since nothing weird happens to the fields $\chi_1$ and $\chi_2$ in the $\gamma \to 0$ limit. We use this example only because it's the simplest case we know of in which logs can arise. There are other examples, see for example \cite{Cardy:2013rqg}, where the two point functions we start from diverge for some value of $\gamma$, and in order to have finite observables we need to construct combinations such as $A$ and $B$. In those cases, logs are forced upon us.} 
It's good to mention that we started with $\chi_1$ having negative norm, while $\chi_2$ has positive norm. This was necessary in terms of continuity in $\gamma$, since, as we already mentioned, by computing the Gram matrix of the logarithmic multiplet $(A_\chi,B_\chi)$, we can see that we have a negative and a positive norm state \cite{Hogervorst:2016itc}.

\subsubsection{Staggered modules}

We've given a simple example of logCFT, but this is not quite what we want yet. If we look at some correlator in this theory where the fields $A_\chi$ and $B_\chi$, as defined in \eqref{eq:ABordinary}, are exchanged, then we expect the conformal block to behave like
\beq
G_{\Delta}\sim |z|^\Delta(1+\alpha \log |z|^2+O(z,\bz))\,,
\eeq
and we see logs appearing at level zero. 
In the $O(n)$ model we have something slightly different. For example, the blocks of current operators have logs only from level one, while Virasoro blocks of operator $\calO_{2,k/2}$ have logs starting at level $k$.

We need a more complicated structure: staggered modules \cite{Gaberdiel:1996kx,Rohsiepe:1996qj}. Here we have a primary operator $\psi_1$ with dimension $\Delta$ which is not logarithmic, but the mixing happens between $B_k$, a level $k$ zero norm descendant of some primary field $\psi_1$, and some unusual operator $A_k$, which is neither a primary nor a descendant. Schematically, ignoring spacetime indices, we have
\beqg
B_{k} \sim \pd^k \psi_1\qquad K B_k =0\\
K^k A_k \sim \psi_1 \qquad A_k \neq \pd^k \psi_1\\
D \begin{pmatrix}
	A_k \\ B_k
\end{pmatrix}=\begin{pmatrix}
	\Delta+k & 1\\
	0 & \Delta+k
\end{pmatrix}  
\begin{pmatrix}
	A_k \\ B_k \end{pmatrix}
\eeqg
where $K$ is the generator of special conformal transformations.
In a four point function where we exchange the operator $\psi_1$, we will find logarithms appearing from level $k$ onwards.\footnote{This, along the property that $A$ cannot be a primary, can be seen for example from the OPE of two fields that exchange $\psi_1$. We will work out the first terms of the $J \cdot \veps$ OPE in section \ref{sec:currents}.}

As an example, we will show how to obtain a level 2 staggered module as a limit of ordinary CFTs. Suppose that we have a theory with two different primary operators, $\psi_1$ and $\psi_2$. We take their dimension to be $[\psi_1]=\Delta+\gamma$ and $[\psi_2]=\Delta+2$. For $\gamma \to 0$ the dimensions of the two operators differ by an integer, with the descendant $\Box \psi_1$ having the same dimension as $\psi_2$. Can they form some sort of logarithmic multiplet? We will see that the answer is positive if $\Delta$ is the dimension of a free field, so that the norm of $\Box \psi_1$ is zero.

The two fields we start with have two point functions
\begin{equation}
	\braket{\psi_1(x)\psi_1(0)}=\frac{-1}{|x|^{2(\Delta+\gamma)}}\qquad \braket{\psi_2(x)\psi_2(0)}=\frac{1}{|x|^{2(\Delta+2)}}
\end{equation}
and the two point function of $\Box \psi_1$ is
\beq
	\braket{\Box \psi_1(x) \Box\! \psi_1(0)}=-\frac{\mathcal{N}_\Box^2}{|x|^{2(\Delta+\gamma+2)}}\,,
\eeq
with
\beq
\mathcal{N}_\Box^2= 4 (\gamma +\Delta ) (\gamma +\Delta +1) (d-2 (\gamma +\Delta +1)) (d-2 (\gamma +\Delta +2))\,.
\eeq
Now let's build the combination
\beqg
A_\psi=\frac{1}{\sqrt{\gamma}}(\psi_2 + \mathcal{N}_\Box^{-1} \Box \psi_1) \\
B_\psi= - \sqrt{\gamma} \psi_2
\eeqg
and, in the $\gamma \to 0$ limit, we have
\beqg
D \begin{pmatrix}
	A_\psi \\ B_\psi
\end{pmatrix}=\begin{pmatrix}
\Delta+2 & 1\\
0 & \Delta+2
\end{pmatrix}  \begin{pmatrix}
A_\psi \\ B_\psi
\end{pmatrix}\\
\braket{A_\psi(x)A_\psi(0)}=\frac{\log |x|^2+O(1)}{|x|^{2(\Delta+2)}}\\
\braket{B_\psi(x)A_\psi(0)}=-\frac{1}{|x|^{2(\Delta+2)}}\\
\braket{B_\psi(x)B_\psi(0)}=0\,.
\eeqg
The $O(1)$ term in the two point function of $A_\psi$ can always be modified by redefining $A_\psi \to A_\psi+\alpha B_\psi$. This ambiguity in defining eigenvectors is usual when we have Jordan blocks, and is always present in logarithmic multiplets.

Everything seems to work nicely so far. However, we also need to check that not only the dilatation operator, but the rest of the conformal generators act nicely on $A_\psi$ and $B_\psi$ as well. Let's look at the generator of special conformal transformation. It's clear that $K^\mu B=0$, but
\beq
K^\mu A_\psi = \frac{\mathcal{N}_\Box^{-1}}{\sqrt{\gamma}} (4 (\Delta+\gamma+1)-2d) \pd^\mu \psi_1 \,.
\eeq

For generic values of $\Delta$, this diverges in the $\gamma \to 0$ limit. However, everything can work out nicely if we choose $\Delta=\frac{d}{2}-1+O(\gamma)$, meaning that in the $\gamma \to 0$ limit, $\psi_1$ is a free field, and its descendant $\Box \psi_1$ has zero-norm (but is not null, as we will see promptly).

So let's choose $\Delta=\frac{d}{2}-1$, and we find that the action of $K^\mu$ on $A$ is finite
\beq
K^\mu A_\psi = \frac{2}{\sqrt{d(d-2)}} \pd^\mu \psi_1 \,.
\eeq
Everything is well defined in this case. Apart from the primary field $\psi_1$, we end up with two fields $A_\psi$ and $B_\psi$ who form a log multiplet. It's clear that $B_\psi$ is a primary; it is also a descendant, since\footnote{We could have  also defined directly $B_{\psi} \sim \Box \psi_1$. The difference between this definition and $B_\psi \sim \sqrt{\gamma}  \psi_2$ is of order $\gamma A$ and vanishes in the $\gamma \to 0$ limit.}
\beq
\Box \psi_1 \sim \gamma A_\psi + B_\psi \xrightarrow[\gamma \to 0]{} B_\psi \,.
\eeq
In unitary CFTs, an operator that is both a primary and a descendant is a null operator, meaning that all of its correlators are zero. This is different here, where $B_\psi$ has norm zero, but is not null.
On the other hand, $A_\psi$ is neither a primary, since $K^\mu A_\psi \ne 0$, nor a descendant, since we cannot write it as the derivative on any other operator of our theory.

This kind of structure is called a staggered module. Since the logarithmic mixing begins at level two (it's a level 2 staggered module), if we look at a four point function where operator $\psi_1$ is exchanged, we will find logarithms at level two. A very similar structure can be found in the $\Box^2$ theory in six dimensions \cite{Brust:2016gjy}; a more complicated staggered structure is shown by the triplet model \cite{Gaberdiel:1998ps}.

In the first paper about logarithmic CFTs \cite{Gurarie:1993xq}, staggered modules were considered in the case of a chiral theory, albeit a different structure was proposed. Translated to our example, the author suggests that operator $A_\psi$ is indeed not a primary, but that there is some new primary operator $\psi_n$ such that $K^\mu A_\psi \sim \pd^\mu \psi_n$. In the $O(n)$ model, our picture looks simpler, since we can count the operators appearing in the torus partition function, and, for a given staggered module, we see no trace of an operator playing the role of $\psi_n$.

\begin{figure}
	\centering
	\includegraphics[scale=.4]{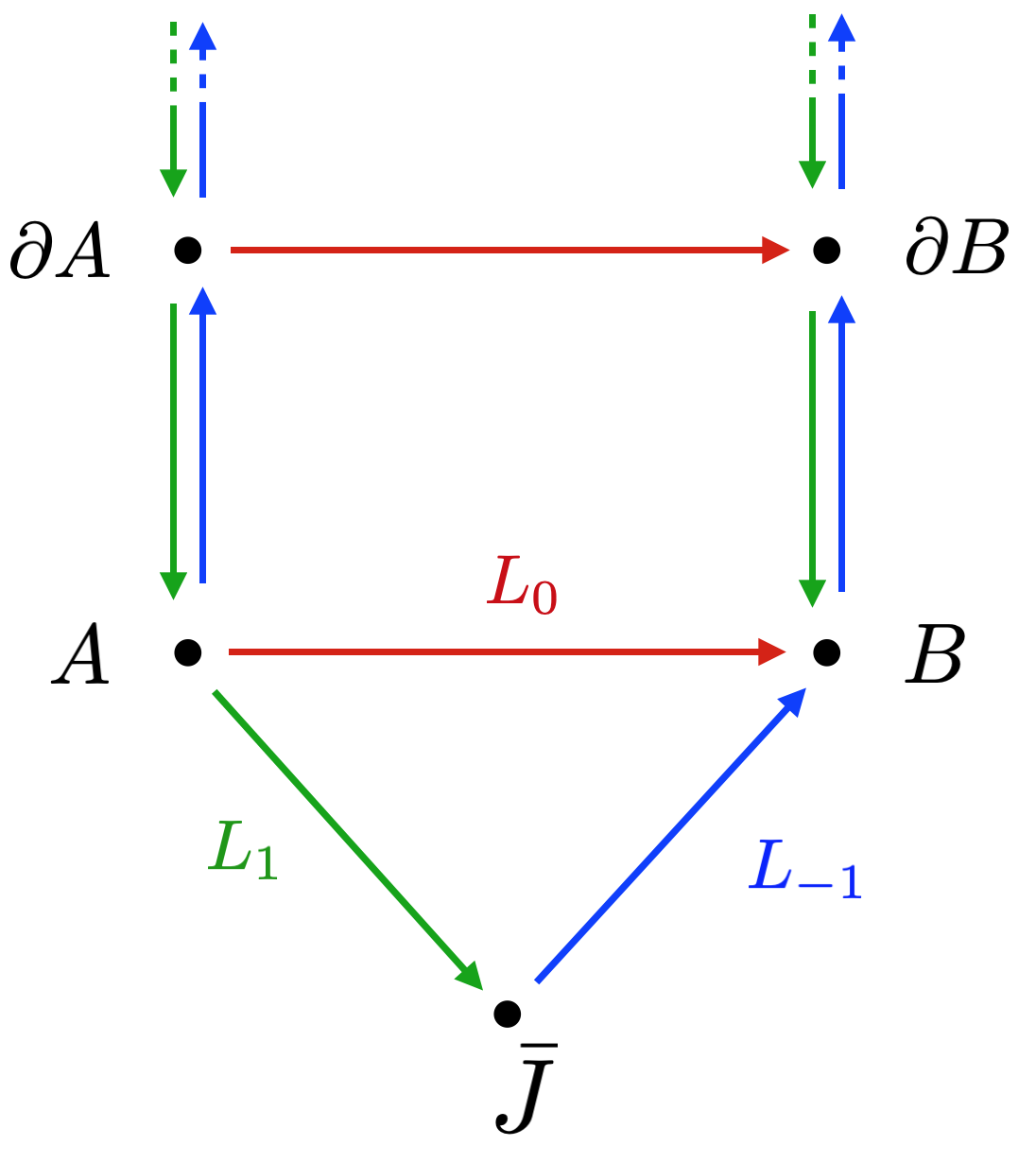}
	\caption{The structure of the staggered module for the currents (colors online). For simplicity we only indicate the action of $L_0$ and $L_{\pm 1}$. $B$ is both a descendant and a primary, while $A$ is neither a descendant nor a primary.}
	\label{fig:staggered}
\end{figure}

In all examples known to us the  presence of a zero norm descendant is necessary for the existence of staggered modules, and in the example above it is $B_\psi \sim \Box \psi_1$.  The requirement that an operator has a zero norm descendant is a non-trivial constraint on when it's possible to have a staggered module. 
As explained in the next subsection (see also Fig.~\ref{fig:staggered}), the $O(n)$ conserved current $J_\mu$ is part of a staggered module: the Virasoro block for the exchange of the current looks like $z+z \bz (a+ b\log |z|^2) +\ldots$, see \eqref{eq:logBlock}, so we expect this logarithmic mixing to appear at level one. The operator $\bpd J$ is indeed zero-norm, but not null. The same happens for the logarithmic operators appearing in the $\veps \cdot \calO_{0,k/2}$ OPE for even $k$. These operators have one of their weight being $h_{k/2,2}$, so, for even $k$, we have a zero-norm descendant at level $k$. Indeed we find that this exchanged operator shows logarithmic mixing starting from level $k$. The condition of having a null descendant is a very constraining condition to find staggered modules, but in the 2d $O(n)$ model the presence of degenerate operators guarantees the existence of zero norm descendants. We comment on the unlikeliness of finding similar structures in higher dimension in section \ref{sec:conclusions}.

\subsubsection{Currents}

After these hopefully pedagogical examples, we go back to the $O(n)$ model. We can now discuss the staggered structure of the current Verma module. We will introduce an artificial deformation of the theory, and then take the limit to the $O(n)$ model to see how logs arise.

The idea would be to change the dimension of the currents $J$ slightly, for example taking it to be $(1+\gamma,\gamma)$. However, one can check that the four point function $\braket{\veps \veps J J}$ would not be crossing symmetric for nonzero $\gamma$. Therefore we assume that we have another scalar primary operator $\zeta$, with weights $(\gamma, \gamma)$ and negative unit norm, of which $J_\zeta$ is a descendant. In the $\gamma \to 0$ limit, we can rescale operators appropriately so that we obtain what we were looking for and $J_\zeta\to J$. We also assume that we have a marginal operator $W$, which will mix with $\pd \bpd \zeta$. 
Let's define
\beqa
J_\zeta&=\frac{1}{\sqrt{2\gamma}}\pd \zeta \\
A_\zeta&=\frac{1}{\sqrt{2 \gamma}}\left( W + \left(\frac{1}{2\gamma}-\frac 12 \right) \pd \bpd \zeta \right)\\
B_\zeta&=-\sqrt{\frac{\gamma}{2}} W\,.
\eeqa
In the $\gamma \to 0$ limit the two point functions of these operators are
\beqa
\braket{J_\zeta(z)J_\zeta(0)}&=-\frac{1}{z^2}\\
\braket{A_\zeta(z)A_\zeta(0)}&=-\frac{\log(z\bz)}{(z\bz)^2}\\
\braket{B_\zeta(z)A_\zeta(0)}&=\frac{1}{2}\frac{1}{(z\bz)^2}\\
\braket{B_\zeta(z)B_\zeta(0)}&=0
\eeqa
and the action of the Virasoro algebra is
\beqg
L_0 \begin{pmatrix}	A_\zeta \\ B_\zeta \end{pmatrix}
=\begin{pmatrix} 1 & 1\\	0 & 1 \end{pmatrix} 
\begin{pmatrix}	A_\zeta \\ B_\zeta \end{pmatrix}\\
L_1 J_\zeta= \bar{L}_1 J_\zeta=0\\
L_1 B_\zeta=0\\
L_{-1} \bJ_\zeta=2 B_\zeta\\
L_{1} A_\zeta= \bJ_\zeta\,.
\eeqg
We see that the operator $\zeta$ effectively decouples from the action of the algebra, and $J_\zeta$ becomes a primary for $\gamma \to 0$. This limit helped us getting some intuition on the structure of our logarithmic multiplets, but we shouldn't take it too seriously. One reason, already mentioned earlier, is that we don't know if there exist a consistent CFT (crossing symmetric and modular invariant) for every value of $\gamma$. A second reason is that we could have taken this limit differently, and some things would have changed. For example, if we assumed that operator $\zeta$ has dimension $(1+2\gamma,1+2\gamma)$, we would have obtained a similar structure but with different coefficients in the relation between operators. We will do everything more carefully, without considering any $\gamma \to 0$ limit, assuming that $L_{-1}\bJ \sim B$ and $L_{1} A \sim \bJ$, but without making any assumption about the coefficients. These coefficients will be determined from necessary properties of correlation functions.

\subsubsection{$\calO_{2,k/2}$ operators} 
Here a similar procedure can shed light on the logarithmic structure of operators $\calO_{2,k/2}$, which appear in the $\veps \cdot \calO_{0,k/2}$ OPE. We will sketch the limit to take in the case of $k=2$. Rather than introducing some new parameter $\gamma$, it's convenient to take $k$ to be some continuous parameter and then take the limit $k\to2$. All the caveats from before still hold: our theory does not contain operators with real values of $k$, so we use this limit again just as a way of building intuition about the logarithmic structure.

Operator $\calO_{2,k/2}$ has dimension
\begin{equation}
	[\calO_{2,k/2}]=(h_{k/2,-2},h_{k/2,2})\,,
\end{equation}
We see that, in the $k \to 2$ limit, we expect a zero-norm descendant at level two. We cannot continue this operator to generic values of $k$ because it would have non-integer spin. Let's assume that we have some scalar primary $\xi$ with dimension $(h_{k/2,2},h_{k/2,2})$, so that we have a level two descendant which becomes zero-norm for $k\to 2$, and another scalar primary $\eta$, with dimension $(h_{1,-2},h_{1,-2})$ and negative norm. We can define our operators
\beqg
\calO_{2,1}\sim \frac{1}{\sqrt{k-2}}\left( L_{-2}-\alpha L_{-1}^2\right) \xi\\
A'_k\sim \frac{1}{\sqrt{k-2}}\left[\eta +\beta  (L_{-2}-\alpha L_{-1}^2)(\bar L_{-2}-\alpha \bar L_{-1}^2)\xi \right]\\
B'_k\sim \sqrt{k-2} \eta
\eeqg
where $\alpha=\frac{3}{2 (2 h_{k/2,2}+1)}$ is chosen so that $\calO_{2,1}$ is a quasi primary for every $k$ and $\beta \sim \frac{1}{k-2}$. When taking the $k \to 2$ limit we find the same sort of two point functions for $A'_k$ and $B'_k$ as we found in the previous section and we have
\beqg
L_{n>0} \calO_{2,1} = \bar L_{n>0} \calO_{2,1} = 0\\
	L_2 A'_2 \sim \bar \calO_{2,1} \qquad L_1 A'_2 =0\\(\bar L_{-2}-\alpha \bar L_{-1}^2) \calO_{2,1} \sim B'_2 \qquad L_{n>0}B'_2=0
\eeqg
Again we have an operator $B'$ which is both a primary and a descendant, and $A'$, which is neither.

\subsection{Structure of the currents} \label{sec:currents}

We will assume that the previous exercise gave us the correct structure of the current Verma module, but we've seen that many of the coefficients that we obtain depend on how we take some fictitious limit. We'll do things more cleanly here. As in the previous section, we have a current $J$ and two dimension two scalars $A$, and $B$,\footnote{We will not use any indices on $A$ and $B$ for the current multiplet.} which mix logarithmically. We normalize them such that
\begin{gather}
\braket{J(z) J(0)}=-\frac{1}{z^2}\\
L_0 A=A+B \label{eq:jordanL0}\\
\braket{A(z)A(0)}=\frac{\log|z|^2}{|z|^4} \label{eq:normA}
\end{gather}
We then assume
\beq
L_{1}A=s \bJ \qquad L_{-1} \bJ = w B \label{eq:ABLsw}\,,
\eeq
where we make no assumption on the coefficients $w$ and $s$.
We will see that $w=-2s$, and we will be able to fix the value of $s$ by looking at some four point function. In our choice of normalization, we have $\pd \bJ=\bpd J=w B$, which fixed the combination of $J$ and $Jb$ conserved to be $\pd \bJ-\bpd J=0$.

As mentioned earlier, a scenario proposed in \cite{Gurarie:1993xq} argued that when we have staggered modules, $L_1 A \sim \bar{J}'$, a different operator from $\bJ$. However, if a $\bar{J}'$ exists, it does not appear in the torus partition function, since by expanding it, we find only $n(n-1)/2$ operators with dimension one and spin one, and this is already $J$. We prefer to avoid a scenario where we have some non-trivial operator who does not appear in the partition function, and, building on the intuition of section \ref{sec:log_limits}, we assume that \eqref{eq:ABLsw} holds.

\subsubsection{Two point functions}
Let's start by fixing the two point functions. It's clear that $\braket{BB}=0$ and $\braket{BJ}=0$, since $B \sim \bpd J$. In order to study two point functions involving $A$, it's good to remember that \eqref{eq:jordanL0} and \eqref{eq:ABLsw} imply an OPE
\beq
T(z)A(0)=\frac{s \bJ(0)}{z^3}+\frac{A(0)+B(0)}{z^2}+\frac{\pd A(0)}{z}+\ldots \,.
\eeq
Let's study the two point function $\braket{A(z_1)X(z_2)}$, where $X=B,J,\bJ$ . The conformal Ward identity, together with the previous OPE, imply
\beq
\sum_{i=1}^2\left[h_i \eps'(z_i)+\eps(z_i)\pd_{z_i} \right] \braket{A(z_1)X(z_2)}+\eps'(z_1)  \braket{B(z_1)X(z_2)}+\frac{s}{2}\eps''(z_1)\braket{\bJ(z_1)X(z_2)}=0 \label{eq:wardId}
\eeq
for $\eps=1,z,z^2$ and something similar for the antiholomorphic part. Solving these equations and imposing $\bpd J = w B$, we find
\beqg
\braket{A(z)J(0)}=\frac{s}{z^2 \bz}\\
\braket{A(z)B(0)}=\frac{s/w}{z^2 \bz^2}\,.
\eeqg
Finally, we look at the $\braket{AA}$ two point function. The conformal Ward identity looks like
\beqg
\sum_{i=1}^2\left[\eps'(z_i)+\eps(z_i)\pd_{z_i} \right] \braket{A(z_1)A(z_2)}+\eps'(z_1)\braket{B(z_1)A(z_2)}+\eps'(z_2)\braket{A(z_1)B(z_2)}\\
+\frac{s}{2}\left(\eps''(z_1)\braket{J(z_1)A(z_2)}+\eps''(z_2)\braket{A(z_1)J(z_2)} \right)=0\,,
\eeqg
and we find
\begin{equation}
	\braket{A(z)A(0)}=-\frac{2s}{w}\frac{\log |z|^2}{|z|^4}\,,
\end{equation}
where we have chosen $A$ so that we don't have any $O(1)$ term in the numerator. We set  $w=-2s$ to have $A$ normalized as in \eqref{eq:normA}.\footnote{The parameter in front of the two point function $\braket{AB}$ is called indecomposability parameter in the literature \cite{Gurarie:2004ce,Vasseur:2011fi}. This parameter has a unique value, but there are different convention in the literature. For example, our convention looks different than that of \cite{Vasseur:2011fi}, because we normalize $\braket{JJ}=-1$. } Finally, denoting $\phi=(A,B)$ the operators of the log multiplet, we have
\beq
\braket{\phi_i(z)\phi_j(0)}=\frac{1}{|z|^4} \begin{pmatrix}
	\log |z|^2 & -\frac{1}{2}\\
	-\frac{1}{2} & 0 
\end{pmatrix} \label{eq:canonical}
\eeq
We are not able to fix the value of $s$ by looking just at two point functions. We will be able to fix it by looking at, for example, the $\braket{JJ\veps\veps}$ four point function, and we will see that it is 
\beq
s= \frac{1}{\sqrt{2m}}\,.
\eeq

We mentioned earlier that one can always redefine $A \to A+\lambda B$, thus changing the value of any correlator where $A$ is inserted. From now on, we will work in the canonical form for which the two point function of $A$ looks like \eqref{eq:canonical}, which removes any ambiguity in the $n$-point functions of $A$.

\subsubsection{Higher point functions and the $J \cdot \veps$ OPE}
We can study the three point functions with insertions of $A$ and $B$ operators. The resulting equations are a bit cumbersome and we put them in appendix \ref{sec:log3pfs}. For the three point functions $\braket{\phi_i \phi_j \veps}$ we have three OPE coefficients, one being $\lambda_{J J \veps}$ and two new ones, $\lambda_{AJ\veps}$ and $\lambda_{AA\veps}$. In principle all of these can be fixed by crossing.

We are ready to work out what the $J \cdot \veps$ OPE looks like. Once we have this, we will insert the OPE in the $\braket{J J \veps \veps}$ four point function, and we'll be able to fix $s$. The first terms of the OPE are\footnote{Notice that acting with $L_0$ or $L_1$ on both sides of the OPE shows that terms such as $\log |z| A$ are not allowed and $A$ is not a primary.}
\begin{equation}
	J(z)\veps(0)=\frac{1}{z^{h_\veps+1}\bz^{h_\veps}} \left[-\lambda_{J J \veps} z J-\lambda_{J \bJ \veps} \bz \bJ +z \bz \left( c_A A+c_B B+c_{log}\log z \bz  B \right) +\ldots \right](0)\,.
\end{equation}
We can fix the coefficients $c$'s by looking at three point functions.
We can compare this OPE to $\braket{J J \veps}$ in the limit of one $J$ being close to $\veps$, and we find
\beq
c_A=\frac{1}{s}\lambda_{JJ \veps}h_\veps\,.
\eeq
Doing the same with the $\braket{A J \veps}$ three point function \eqref{eq:AJeps}, we get
\beqg
c_{log}=c_A\\
c_B=-2\lambda_{AJ\veps}-4 s \lambda_{JJ\veps} h_\veps\,.
\eeqg

Let's take now the $\braket{\veps(0) J(z) J(1) \veps(\infty)}$ four point function. In a small $z,\bz$ limit, we have (this corresponds to the $z \to 1$, $\bz \to 1$ limit in equation \eqref{eq:JJtchan})
\beq
\braket{\veps(0) J(z) J(1) \veps(\infty)}=\frac{1}{z^{h_\veps+1}\bz^{h_\veps}}\left( f_{1,0} z+ f_{0,1}\bz+ f_{1,1} z \bz +g_{1,1}z \bz \log(z \bz)+\ldots \right) 
\eeq
By again inserting the $\veps \cdot J$ OPE in the four point function, apart from the expected $f_{1,0}=-\lambda_{JJ\veps}^2$ and $f_{0,1}=-\lambda_{J\bJ\veps}^2$, we find the following relations 
\beqg
f_{1,1}=-2 h_\veps \lambda_{J J \veps} \left( \frac{\lambda_{AJ\veps}}{s}+h_\veps \lambda_{JJ\veps}\right) \\
g_{1,1}=\frac{h_\veps^2 \lambda_{JJ\veps}^2}{2 s^2}\,.
\eeqg
From these relations we get\footnote{In order to fix $s$ we do not need to impose crossing of the four point function, which we will instead need for fixing $\lambda_{AJ\veps}$. The reason behind this is that, from \eqref{eq:JJtchan}, logarithms arise only from $\widetilde{\calV}_0$ and $\widetilde{\bar \calV}_0$. It's enough to solve the BPZ differential equation and check that $\widetilde{\bar \calV}_0=1-\frac{(m-1)^2 m}{(m+1)^2} \bz \log \bz+\ldots$ to conclude that $g_{1,1}/f_{1,0}=-\frac{(m-1)^2 m}{(m+1)^2}$. Notice that we could have also used a different four point function to fix $s$.}
\beq
s=\pm h_\veps \lambda_{JJ \veps}\frac{1}{\sqrt{2 g_{1,1}}} = \pm \frac{1}{\sqrt{2m}}, \label{eq:s}
\eeq
exactly, and the OPE coefficient $\lambda_{AJ\veps}$, of which we keep an expression only to first order in $1/m$ 
\beq
\lambda_{AJ\veps}=\mp \sqrt{\frac{m}{6}}+O(1)\,, \label{eq:OPEA}
\eeq
where $A$ is defined so that it's two point function is canonical \eqref{eq:canonical}. The ambiguity in sign corresponds to the possibility of redefining $J \to -J$ in \eqref{eq:ABLsw}, so we will just choose the upper sign for both $s$ and $\lambda_{AJ \veps}$.

We will not look explicitly at operators $\calO_{2,k/2}$ in this work, but the procedure to fix the logarithmic structure and OPE coefficients is straightforwardly applicable to them as well.

\subsection{Validity of the BPZ differential equations}
\label{sec:BPZ_validity}
Our approach to obtaining OPE coefficients relies on the fact that an operator has a null descendant, and therefore correlation functions of this operator satisfy a BPZ differential equation. However, these differential equations are sometimes modified in logarithmic CFTs \cite{Flohr:2001zs}. One then might wonder if our approach is self consistent, or if our differential equation \eqref{eq:BPZ4pf} needs to be somehow modified. We will show that everything is fine for the correlators we considered in section \ref{sec:crossing}.

We will first give an example a four point function, which we have not considered in section \ref{sec:crossing}, where differential equations would be indeed modified, i.e. the case of a correlator where the field $A$ is an external operator.
Notice that $A$ is not a descendant, so we cannot find its correlators by acting with some differential operator on n-point functions of primaries. The OPE $T \cdot A$ is
\begin{equation}
T(z)A(0)=\sum_n \frac{L_n A(0)}{z^{2+n}}=\frac{s \bJ(0)}{z^3}+\frac{A(0)+B(0)}{z^2}+ \frac{\pd A(0)}{z}+\ldots\,.
\end{equation}
The extra term with $\bJ$ is the least unusual one, and is there because $A$ is not a primary; the term with $B$ instead is due to the logarithmic nature of $A$, $(L_0-1)A=B$. Let us consider the correlator $\braket{\veps A\calO_2 \calO_3 }$, and let us act with one $L_{-n}$ on $\veps$. We have
\beqg
\braket{L_{-n}\veps(z) A(z_1) \calO_2(z_2)\calO_3 (z_3)}=\calL_{-n}\braket{\veps(z) A(z_1) \calO_2(z_2)\calO_3 (z_3)}
+\frac{n-1}{(z_j-z)^n}\braket{\veps(z) B(z_1) \calO_2(z_2)\calO_3 (z_3)}-\\-\frac{s}{2}\frac{n (n-1)}{(z_j-z)^{n+1}}\braket{\veps(z) \bJ(z_1) \calO_2(z_2)\calO_3 (z_3)}
\eeqg
where $\calL_n$ is defined as \eqref{eq:Ldiff}, and $h_1=1$. Using this equation and the fact that $\veps$ has a null descendant, we can get an differential equation for $\braket{\veps A\calO_2 \calO_3 }$, but this will be inhomogeneous and involve the correlators $\braket{\veps \bJ \calO_2 \calO_3 }$ and $\braket{\veps B \calO_2 \calO_3 }$ as well. These extra terms modify the BPZ differential equation.

If we consider instead a correlator with no insertions of $A$, such as $\braket{JJ\veps \veps}$, then the differential equation \eqref{eq:BPZ4pf} is unchanged. This is because if we act with $L_{-n}$ on $J$, we do not find any non-diagonal action, hence the $T \cdot J$ OPE looks ordinary and has no unusual terms.\footnote{As mentioned earlier, $J$ and $B$ are part of an invariant submodule under the Virasoro action.} In this work, we limit ourselves to study crossing equations for correlators without insertions of $A$ or other logarithmic operators such as $\calO_{2,k/2}$, meaning that, for example, we will not be able to fix from crossing the OPE coefficient $\lambda_{AA\veps}$. We will still compute this coefficient at leading order by studying the $n\to2$ limit.

\section{The critical to low-T flow to first order}
\label{sec:flow}
Now that we know both anomalous dimensions and OPE coefficients of some leading operators we can come back to studying the RG flow from 
 critical to low-T fixed point in conformal perturbation theory. We follow the general recipe of \cite{Cardy_book} and work at leading order in $\sqrt{n-2}\sim\frac{1}{m}$. As we already discussed, the only relevant singlet in the UV theory is the operator $\veps$, which must be responsible for triggering the RG flow to the low-T fixed point.
It has dimension
\beq
\Delta_\veps=2\mp \frac{4}{m}+\ldots\, \label{eq:dim_eps}
\eeq
where, for the rest of the section, the upper (lower) sign refers to the critical (low-T) fixed point.
We perturb the critical theory by
\beq
S_{crit}+g\int d^2 x \veps(x)
\eeq
which gives a beta function
\beq
\beta(g)=(\Delta_\veps^{{crit}}-2) g+\pi \lambda_{\veps \veps \veps} g^2+\ldots\,.
\eeq
where by $\lambda_{\veps \veps \veps}$ we mean the OPE coefficients taken at $m=\infty$ ($n=2$).
The IR fixed point is at $g^*=\frac{2-\Delta_\veps^{{crit}}}{\pi \lambda_{\veps \veps \veps}}$, and it can be seen that the IR dimension of the energy field is
\beq
[\veps]_{\rm IR} = 2+\beta'(g^*)=2+\frac{4}{m}+\ldots
\eeq
This, as usual in one loop conformal perturbation theory, is a trivial result, as it does not depend on the value of $\lambda_{ \veps \veps \veps}$. 

Now let's consider other operators. %
For a non-logarithmic operator $\phi$ its coupling $u_\phi$ has a beta function
\beq
\beta_\phi(u_\phi)=(\Delta_\phi-2)u_\phi+2\pi \lambda_{\phi \phi \veps} u_\phi g
\eeq
where we have neglected higher order terms as well as terms which vanish at the fixed point $g=g^*$ and $u_\phi=0$. The IR dimension for $\phi$ is
\beq
[\phi]_{\rm IR}=2+\left. \pd_{u_\phi} \beta_\phi\right|_{\substack{g=g^*\\u_\phi=0}}=[\phi]_{\rm UV}+8\frac{\lambda_{ \phi \phi \veps}}{\lambda_{ \veps \veps \veps}}\frac{1}{m} \label{eq:andim}
\eeq

Next we check the operators $\calO_{0,k/2}$, which correspond to the spin operator for $k=1$. Their dimension is
\beq
\left[\calO_{0,k/2}\right]=\frac{k^2}{8}\pm \frac{k^2}{8 m}+\ldots
\eeq
This and \eqref{eq:andim} is consistent with the result found in section \ref{sec:O0k} for $n\to2$,
\beq
\label{lambdaOk}
\lambda_{\calO_{0,k/2} \calO_{0,k/2}\veps}=-\frac{k^2}{8\sqrt{3}} \,.
\eeq
This provides a check for our bootstrap procedure.
In this calculation we assumed that the operators in question do not mix with any other fields under RG flow. While the spin has no field with which it may mix, this is not true for general $k$. For example, operator $\calO_{0,2}$ has dimension close to two, and could mix with the marginal operators $A$ and $B$ at the leading order in CPT. In section \ref{sec:marginal} we will see that among operators $\calO_{0,2}$ there cannot be an adjoint of $O(n)$ and consequently they cannot mix with $A$ or $B$.
Formula \ref{lambdaOk} strongly suggests that none of the the $\calO_{0,k/2}$ mix with other operators, since mixing in general would spoil the agreement of conformal perturbation theory with the spectrum of the low-T theory. This can be checked explicitly for low values of $k$ by computing corresponding OPE coefficients from bootstrap. In the next subsection we will see that operator mixing is in fact present in other sectors.

An expected result is that the current does not renormalize. This is easy to see since $\lambda_{JJ \veps} =0$ at $n=2$.

\subsection{Operators mixing} \label{sec:mixing}
There are cases in which, instead, we need to consider mixing between operators. The simplest example we found is the operator $\calO_{2,1}$. This operator has a zero norm descendant at level 2 which is a part of a logarithmic multiplet. However, the rules of one loop conformal perturbation theory for $\calO_{2,1}$ itself do not change.

We have
\beq
\left[\calO_{2,1}\right]=\frac{5}{2}\mp \frac{3}{2 m}+\ldots\,,
\eeq
while the $n=2$ OPE coefficient is
\beq
\lambda_{\calO_{2,1}\calO_{2,1}\veps}=\frac{3\sqrt{3}}{2}\,, 	
\eeq
which leads to a mismatch.
Consequently, we need to look at other operators with which $\calO_{2,1}$ might mix. The only primary around with the right spin and dimension is operator $\calO_{1,2}$, who has dimension $\frac{5}{2}\pm \frac{3}{2m}+\ldots$. However, using crossing equations for the correlator $ \braket{\calO_{1,2} \calO_{2,1} \veps \veps}$, it can be checked that $\braket{\calO_{1,2} \calO_{2,1} \veps}=0$. 

There is, however, another quasi-primary around, the level two descendant of $\calO_{0,1}$ 
\beq
\Omega = (L_{-2}-\alpha L_{-1}^2 )\calO_{0,1}\,,
\eeq
where $\alpha$ is chosen so that it's a quasiprimary
\beq
\alpha=\frac{3}{2(1+2h_{{m-1},{m+1}})}\,.
\eeq
This descendant becomes zero norm as $m \to \infty $, as its norm is given by
\beq
\braket{\Omega \Omega}=- \frac{8}{3m^2}+\ldots\,,
\eeq
In the same manner, it can be checked that the three point functions involving $\Omega$ go to zero when $m$ goes to $\infty$. Note that for $n\lesssim2$ this operator has a negative norm.

To determine the three-point functions we first use crossing equations for the correlator $\braket{\calO_{0,1}\calO_{0,1}\veps \veps}$, where $\calO_{2,1}$ is exchanged in the t-channel. This allows us to determine
\beq
\lambda_{\calO_{0,1}\calO_{2,1} \veps}=\pm \sqrt{\frac{2}{3}}+\ldots\,.
\eeq
The sign is unimportant, so we just choose the OPE coefficient to be positive. Next we define $\widetilde{\Omega} = i \frac{3}{2\sqrt{2}}m\Omega$, so that the two point function of $\widetilde{\Omega}$ has coefficient one.\footnote{The reason we are doing this rescaling is that we are used to do conformal perturbation theory for operators normalized so that they have a two point function with coefficient one. A reader bothered by this introduction of $i$ could rework the rules of perturbation theory with a different normalization and would arrive to the same result.} Using the operators \eqref{eq:Ldiff} on the three point function it's easy to see that,
\beqg
\lambda_{\widetilde{\Omega}\calO_{2,1} \veps}=-i\sqrt{6}\lambda_{\calO_{0,1}\calO_{2,1} \veps}=\pm 2 i\\
\lambda_{\widetilde{\Omega} \widetilde{\Omega}\veps}=7 \lambda_{\calO_{0,1}\calO_{0,1} \veps}=-\frac{7}{2\sqrt{3}}\,.
\eeqg
Indeed we have a rather peculiar effect of mixing between the two different Verma modules.
Notice that the OPE coefficients of $\widetilde{\Omega}$ operator now are finite in the $m \to \infty$ limit.

Now we consider the beta function for the two operators
\beq
\beta_i=(\Delta_i-2)u_i+2\pi \lambda_{i j \veps} u_{j} g
\eeq
with the indices $i,j$ taking values $\calO_{2,1}$ and $\widetilde{\Omega}$. At the fixed point we have
\beq
2\delta_{ij} + \pd_{u_i}\beta_j=
\begin{pmatrix}
	\frac{5}{2}-\frac{3}{2m}& \\
	& \frac{5}{2}+\frac{1}{2m}
\end{pmatrix}
+\frac{1}{m}
\begin{pmatrix}
	9 & \pm i 4\sqrt{3}\\
	\pm i 4\sqrt{3}	& -7
\end{pmatrix}
\eeq
The eigenvalues of this matrix are the dimensions of our IR operators, and they are $\frac{5}{2}-\frac{1}{2m}$ and $\frac{5}{2}+\frac{3}{2m}$; this agrees with the low temperature spectrum. 

\subsection{Logarithmic operators}

We've seen, by checking explicitly at one loop, that the current does not renormalize; it is natural to wonder what happens to the logarithmic multiplet formed by operators $A$ and $B$. We still expect a descendant of the current to have dimension 2 and spin 0 in the IR, but it's not clear what the fate of the other operator is.

If we consider two non logarithmic operators that have the same dimension in the UV, we expect them generically to have different dimensions in the IR. This is because we expect these operators to mix, and when we diagonalize the beta functions we expect to find eigenvalue repulsion (unless there is a symmetry reason for this not to happen).
So intuitively, we would expect a logarithmic multiplet to be unstable against perturbations, as the three point function $\braket{AB\veps}$ is non zero \eqref{eq:AJeps},\footnote{Notice that while we might redefine $A \to A+\lambda B$, the three point function \eqref{eq:AJeps} cannot be made zero.} so that the two operators mix. However, we can see from the low-T partition function that operators $A$ and $B$ are marginal in the IR as well, and by studying correlators such as $\braket{JJ\veps\veps}$ we can see that the logarithmic structure survives.

Therefore there needs to be some more subtle way in which the logarithmic structure is preserved. To check this explicitly, we would need to derive the rules of logarithmic conformal perturbation theory. We leave this question for future work.
Here we only mention quickly which kind of divergences we expect to see. Let's consider the first order correction to the two point functions of the logarithmic multiplet, and use an analytic continuation in $1/m$ in order to make the integral of the three point function finite. When integrating an ordinary three point function we would find generically divergences linear in $m$. 
However, OPE coefficients have their own $m$-dependence, which changes the structure of divergences (see Appendix \ref{sec:log3pfs} for the explicit form of the three point functions). In particular, the integrated three point function $\braket{BB\veps}$ will give us no divergences, since the OPE coefficient is order $1/m$. The integral of $\braket{A B \veps}$ will give divergences of order $m$; the $m^2$ divergence arising from the logarithmic term is made softer by its coefficient. In the same way, we expect divergences of order $m^2$ from $\braket{A A \veps}$ since the $m^3$ divergence arising from the $\log^2$ term has a $1/m$ coefficient. We see that logarithmic terms are prone to produce higher order divergences and it remains non trivial to see why both $A$ and $B$ remain marginal logarithmic operators in the IR fixed point; however, in our case, the highest possible divergence is softened by the specific behavior of the OPE coefficients.

\section{The $n \to 2$ limit}
\label{sec:n->2}
We would like to make contact between what we've studied so far and the free boson formulation of the $O(2)$ model. The $O(2)$ model describes the celebrated BKT phase transition \cite{Kosterlitz:1973xp} and the corresponding CFT can be described by a compactified free boson with radius $R=1/\sqrt{2}$ \cite{Dijkgraaf:1987vp}. This theory is unitary and has an enhanced symmetry $O(2)\times O(2)$, while as we have seen the $O(n)$ model for generic $n$ is non-unitary, logarithmic, and has no symmetry enhancement. In this section we will study how negative norm states drop out of the theory and how logarithms drop out of correlation functions as $n \to 2$.

Let's start by identifying the most important $O(2)$ operators in the free boson formulation. The theory has action
\beq
S=\frac{1}{8\pi}\int d^2 x (\pd_\mu \phi)^2\,.
\eeq
The free field is the sum of a holomorphic and antiholomorphic component
\beq
\phi(z,\bz)=\vphi(z)+\bvphi(\bz)\,,
\eeq
and correlation functions can be built using the Wick theorem with the propagator
\beq
\braket{\phi(z,\bz) \phi(0)}=-\log |z|^2\,.
\eeq
Apart from operators which we can build out of derivatives of $\vphi$ and $\bvphi$, we can also build vertex operators
\beqa
V^+_{n,m}&=\sqrt{2}\cos(p \vphi+\bp \bvphi)\\
V^-_{n,m}&=\sqrt{2}\sin(p \vphi+\bp \bvphi)
\eeqa
with dimension $(\frac{p^2}{2},\frac{\bp^2}{2})$, 
where $n,m$ are integers and are related to $p, \bp$ by
\beq \label{eq:pnm}
(p,\bp)=\left( \sqrt{2}n+\frac{1}{2\sqrt{2}}m,\sqrt{2}n-\frac{1}{2\sqrt{2}}m\right) \,,
\eeq
The theory contains other primary operators which schematically look like $\pd^k \vphi V_{n,m}$ if $\sqrt{2}p$ is integer, and similarly for the antiholomorphic part. One of such operators will be important for our discussion below. %

We can immediately identify the currents, since the only dimension one vector operators are $\pd \vphi$ and $\bpd \bvphi$. These currents are conserved as a consequence of the equations of motion $\bpd \pd \phi=0$. The two $O(2)$ symmetries correspond to independent shifts of $\vphi$ and $\bvphi$, as well as the $(\vphi,\bvphi) \to-(\vphi,\bvphi)$ and $(\vphi,\bvphi) \to (\bvphi,\vphi)$ transformations.
We can also immediately identify the spin as the only scalars with dimension $1/8$,  the vertex operators $V^\pm_{0,1}$. It is a bit harder to identify what is the energy operator $\veps$, since there are five marginal operators: $V_{1,0}^\pm$, $V_{0,4}^\pm$ and $L=\pd \vphi \bpd \bvphi$. $L$ is the exactly marginal operator, adding which to the action corresponds to changing the radius of the boson.

At this point, it's important we consider the question of symmetry for continuous $n$. For $n \le 2$, the symmetry group is $O(n)$, while for $n=2$ it's enhanced to $O(2)\times O(2)$ at the critical point. We assume that as $n \to 2$, $O(n) \to O(2)_p \subset O(2)\times O(2)$. The ``physical'' $O(2)_p$ is, in particular, still a symmetry away from the critical point, where we don't expect any symmetry enhancement. Below we will determine what $O(2)_p$ is in the free boson formulation from various consistency requirements, being agnostic to its microscopic nature.
For example, the energy operator needs to be a singlet of $O(n)$ for $n<2$, so we expect it to be a singlet under $O(2)_p$ only, rather than under the full $O(2)\times O(2)$ group. It's clear that  the energy operator cannot indeed be a singlet of $O(2)\times O(2)$, since the only marginal operator which is singlet under this group is $L$; $L$ is an exactly marginal operator, meaning that its three point function $\braket{LLL}$ vanishes, while we know that $\braket{\veps \veps \veps}\neq 0$ at $n=2$.
 
This means that some of the operators $V_{1,0}^\pm$ and $V_{0,4}^\pm$ have to be singlet under $O(2)_p$. Let's focus for the moment on the $U(1)_p$ subgroup, $U(1)_p\subset O(2)_p$. The operators $V_{1,0}$ are singlets only under the single combination of  the two $U(1)$'s that leaves $\vphi +\bvphi$ invariant, while the operators $V_{0,4}$ are singlet under the combination that leaves $\vphi -\bvphi$ invariant. The requirement of either of these two operators being a singlet under $O(2)_p$, means that $U(1)_p$ can only be one of the two mentioned above. To decide among which one, we should remember that the spin operator is $V_{0,1}$ and it must be charged under $U(1)_p$.
This allows us to identify
 \beq
 U(1)_p: \vphi \to \vphi+a,\bvphi \to \bvphi-a \,.
 \eeq
 It follows that the energy operator in this formulation must be a linear combination of $L$ and $V_{1,0}^\pm$. We identify it by requiring $\lambda_{\veps \veps \veps}=4/\sqrt{3}$. The relevant OPE coefficients for the $O(2) $ model are\footnote{Formula (2.14) in \cite{Dijkgraaf:1987vp} is missing a minus sign.}
\beq
\lambda_{L V_{n,m}^\pm  V_{n,m}^\pm }
=-\left(2 n^2-\frac{m^2}{8} \right) 
\eeq
and the only vertex operators appearing in the OPE of $V_{m,n}$ and $V_{m',n'}$ is $V_{m+m',n+n'}$ and $V_{m-m',n-n'}$.
Imposing $\lambda_{\veps \veps \veps}=4/\sqrt{3}$ we have
\beq
\veps=\frac{1}{\sqrt{1+b_+^2+b_-^2}}\left(-L +b_+ V_{1,0}^++b_- V_{1,0}^-\right) \qquad \text{with} \quad b_+^2+b_-^2=2\,.
\eeq
Without the loss of generality we make the choice $b_-=0$ and $b_+$ positive (other choices are related by the remaining $O(2)$ rotation and only affect the definition of our physical $\mathbb{Z}_2$), so we identify
\beq
\veps= -\frac{1}{\sqrt{3}}L+\sqrt{\frac{2}{3}}V_{1,0}^+\,.
\eeq
We will need some more information from the $n\less2$ theory to identify the action of the physical $\mathbb{Z}_2$ and we return to this question below.

We can check that, as expected, $\lambda_{\veps \sigma \sigma}=-\frac{1}{8\sqrt{3}}$.
We can  easily identify operators $\calO_{0,k/2}=V_{0,k}$ for $k=2$ and $k=3$ based just on their dimensions. Corresponding OPE coefficients, computed in the free boson formulation, also match nicely our results from section \ref{sec:O0k}, so the $n\to 2$ limit is smooth and simple in this case; however, for higher dimensions the limit is more subtle. The $k=4$ case is discussed in details below.

\subsection{Decoupling of negative norm states}
We've seen that the critical $O(n)$ model is non-unitary for generic values of $n$.
For $n\to 1,2$, however, we need to recover in some way a unitary theory. This would mean that all negative norm  and logarithmic states have to drop out of the theory. We will illustrate here how this process works, and give a few explicit examples that corroborate our scenario. We will consider exclusively the $n\to 2$ limit, but we conjecture that a similar scenario works for the $n \to 1$ limit as well.

What makes this issue tricky is that taking the limit $n \to 1,2$ does not yield the unitary theory at $n=1,2$, as was already emphasized in \cite{Cardy:2013rqg}. In fact the $n \to 1,2$ theory always contains more operators. Operators of the unitary theory form a closed subalgebra, to which the $n \to 1,2$ theories can be consistently truncated.\footnote{Discrete values of $n$ often lead to theories with some special properties. See, for example, \cite{Vasseur:2011fi,saleur1991antiferromagnetic,Jacobsen_2006}.} To illustrate this subtlety, consider an operator $\calO_{1,2}$ that we briefly mentioned in section~\ref{sec:mixing}. Its multiplicity is given by $n(n+1)^2(n-2)/4$, which smoothly goes to zero at $n=2$. At first thought one may be tempted to simply ignore this operator, however, we will see momentarily that this is too quick. Instead, multiplicity $n(n+1)^2(n-2)/4$ does not correspond to the dimension of an irreducible representation and there are several distinct operators with dimension of $\calO_{1,2}$. Some of them are part of the unitary $n=2$ subalgebra, while others decouple. Since the total multiplicity is zero it means that decoupled operators have negative multiplicity, which means that some operators with the same dimension and positive multiplicity additionally decouple at $n=2$. As we saw in \ref{sec:mixing} such operators are indeed present in the theory, and moreover some of them have negative norm. We come back to this particular set of operators in section~\ref{sec:nonlog}, while we begin the discussion of this mechanism in the sector of more important operators, namely with the currents logarithmic multiplet.

\subsubsection{Logarithmic operators and the marginal sector}
\label{sec:marginal}
As we previously discussed, conserved $O(n)$ currents are part of the logarithmic multiplet which necessarily contains negative norm states. Clearly currents themselves must survive in the $n\to2$ limit, while some of the logarithmic operators  $A$ and $B$ must decouple. This appears puzzling because the total number of marginal operators at $n=2$ matches nicely combined multiplicity of $\veps$, $A$ and $B$ and $\calO_{0,2}$:
\beq
1+n(n-1)+\frac{1}{4}n(n-1)^2 (n+2)\,\Big|_{n=2}=5\,.
\eeq
The resolution of this puzzle was outlined just above. Some negative multiplicity operator inside $\calO_{0,2}$ should decouple and cancel the contribution to the partition function of a negative-norm combination of $A$ and $B$. In the course of this, logarithms will disappear from the correlation functions of the currents and they will become holomorphic. Let us see how this happens in some details.

We can combine $A$ and $B$ into the two operators
\beq
\psi_{1}=\frac{\sqrt{m}}{r} B- \frac{r}{\sqrt{m}}A\qquad \psi_{2}=\frac{\sqrt{m}}{r} B+ \frac{r}{\sqrt{m}}A
\eeq
where $r$ is some constant yet to be determined. In the $m \to \infty$ limit, we have the two point functions
\beqg
\braket{\psi_{1}(x) \psi_{1}(0)}=\frac{1}{|x|^4} \qquad \braket{\psi_{2}(x) \psi_{2}(0)}=\frac{-1}{|x|^4} \\
\braket{\psi_{1}(x) \psi_{2}(0)}=0
\eeqg
Therefore, in the $n \to 2$ limit, the two operators $\psi_i$ are orthogonal and their two point functions are simply power laws, as in this basis logarithmic corrections are suppressed by $1/m$. However, $\psi_1$ has positive norm, whereas $\psi_2$ has negative norm. We then expect $\psi_2$ to drop out of the unitary sector of the theory. We can check explicitly whether this happens or not for some three point function in the $n\to 2$ limit. Let's consider $\braket{\psi_i J \veps}$: using \eqref{eq:AJeps} and \eqref{eq:JJe_OPE},\footnote{As usual, we are considering $A$ to be in its canonical form, \eqref{eq:canonical}.} we get 
\beqg
\frac{\braket{B(z_1) J(z_2) \veps(z_3)}}{\llangle B(z_1) J(z_2) \veps(z_3) \rrangle}=\lambda_{B J \veps}=\frac{h_\veps}{2s}\lambda_{JJ \veps}=\sqrt{\frac{2}{3 m}}+O\left( \frac{1}{m}\right) \\
\frac{\braket{A(z_1) J(z_2) \veps(z_3)}}{\llangle A(z_1) J(z_2) \veps(z_3) \rrangle}=\lambda_{A J \veps}+\ldots=-\sqrt{\frac{m}{6}}+O(1)\,
\eeqg
where we've introduced the notation $\llangle \ldots \rrangle$ to indicate the usual coordinate dependence of an ordinary three point function of primaries,
\begin{equation}
\llangle \calO_1(z_1) \calO_2(z_2) \calO_3(z_3) \rrangle = \frac{1}{z_{12}^{h_{123}} z_{23}^{h_{231}} z_{13}^{h_{132}}}\frac{1}{\bz_{12}^{\bar h_{123}} \bz_{23}^{\bar h_{231}} \bz_{13}^{\bar h_{132}}}\,.
\end{equation}
with $h_{ijk}=h_i+h_j-h_k$. Notice that the coordinate dependence of correlation functions of logarithmic operators can be quite complicated (see the explicit formulas of appendix \ref{sec:log3pfs}). In the $m \to \infty$ limit, the leading behavior of $\braket{AJ\veps}$ takes a simpler form and looks like that of an ordinary three point function of primaries.

Combining $A$ and $B$ in their linear combinations $\psi_1$ and $\psi_2$, we find
\beqg
\lambda_{\psi_1 J \veps}=\frac{1}{r}\sqrt{\frac{2}{3}}+\frac{r}{\sqrt{6}}\\
\lambda_{\psi_2 J \veps}=\frac{1}{r}\sqrt{\frac{2}{3}}-\frac{r}{\sqrt{6}}
\eeqg
Requiring $\psi_2$ to decouple from the unitary sector of the theory imposes $r=\pm \sqrt{2}$, and implies $\lambda_{\psi_1 J \veps}=\pm \frac{2}{\sqrt{3}}$. Therefore we have constructed a positive norm operator $\psi_1$ which is part of the unitary sector of the theory, and a negative norm operator $\psi_2$ which decouples from the unitary sector of the theory. From here on, we will choose the plus sign for $r$.

We haven't studied four point functions with $A$ as an external operator since, as explained in section \ref{sec:BPZ_validity}, they satisfy a more complicated inhomogeneous equation, so we have no access to the OPE coefficient $\lambda_{AA\veps}$.
We need it to diverge at most as $m$ in order for $\braket{\psi_i \psi_j \veps}$ to be finite. Let's assume that's indeed how it diverges
\beq
\lambda_{AA\veps}=c_0 m+O(\sqrt{m})\,.
\eeq
 We will now show that the value of $c_0$ needs to satisfy different constraints, and despite having more equations than unknowns, there is one value of $c_0$ that satisfies all the constraints.
Using the formulas in appendix \ref{sec:log3pfs}, we have 
\beqg
\frac{\braket{B(z_1) B(z_2) \veps(z_3)}}{\llangle B(z_1) B(z_2) \veps(z_3) \rrangle}=\frac{2}{\sqrt{3}}\frac{1}{m}+O\left( \frac{1}{\sqrt{m}}\right)\\
\frac{\braket{A(z_1) B(z_2) \veps(z_3)}}{\llangle A(z_1) B(z_2) \veps(z_3) \rrangle}=\frac{\lambda_{AJ \veps}(h_\veps-1)}{2s}+\frac{h_\veps \lambda_{JJ\veps}}{4s^2}+O\left(\frac{1}{\sqrt{m}} \right)= O\left(\frac{1}{\sqrt{m}} \right)\\
\frac{\braket{A(z_1) A(z_2) \veps(z_3)}}{\llangle A(z_1) A(z_2) \veps(z_3) \rrangle}=c_0 m+O(\sqrt{m})\,.
\eeqg
Therefore we have
\beqg \label{eq:psi_n2}
\lambda_{\psi_i \psi_i \veps}=\frac{1}{\sqrt{3}}+2c_0\\
\lambda_{\psi_1 \psi_2 \veps}=\frac{1}{\sqrt{3}}-2c_0\,.
\eeqg
In order to have $\psi_2$ decoupling from the unitary sector of the theory, we need it not to appear in any OPE of operators of the unitary sector. This means that we need $c_0=\frac{1}{2\sqrt{3}}$. The study of the four point function $\braket{AJ \veps \veps}$, for example, would  allow us to study the OPE coefficient $\lambda_{AA\veps}$, thus allowing to get yet another constraint on $c_0$. 

Let us also mention however that if indeed $c_0=\frac{1}{2\sqrt{3}}$, then
\beq
\label{lambdapsi} 
\lambda_{\psi_1 \psi_1 \veps}=\frac{2}{\sqrt{3}}\,.
\eeq 

Next, let us identify the $n\to2$ limit of this operator with an operator in the $O(2)$ theory. For this we need to come back to identification of the physical $\mathbb{Z}_2$ symmetry, a question that we still owe to the reader. Operator $\psi_1$ is in the adjoint representation of $O(n)$ for a generic $n$, hence at $n\to2$ it becomes a pseudoscalar of $O(2)$; this leaves the only possible identification $\psi_1\to V_{1,0}^-$ and requires that $V_{1,0}^-$ transforms under $\mathbb{Z}_2$ non-trivially (while $V_{1,0}^+$ must be a singlet).\footnote{Note that this $\mathbb{Z}_2$ leaves neither electric nor magnetic operators invariant.} This fixes the symmetry to act as $(\vphi,\bvphi)\to(-\vphi,-\bvphi)$. 
Using the formulas collected in appendix \ref{sec:O2}, it is easy to check that \ref{lambdapsi}, as well as another 
 three point function which we also determined at $n\less2$, $\lambda_{\psi_1 J \veps}=\frac{2}{\sqrt{3}}$, is compatible with the identification $\psi_1=V_{1,0}^-$. Let us remark once more that what we're doing here is more than just guesswork: we have several constraints on $c_0$, coming from different three point function, so it's highly non-trivial that there exist a value of $c_0$ which satisfies all the constraints. 

Formulas \eqref{eq:psi_n2} also imply that $\braket{\psi_2 \psi_2 \veps}\neq0$. There is no conflict between this result and our statement that there is a closed subalgebra of positive norm operators: in the OPE of a positive norm and a negative norm operator anything can appear, our statement only concerns the OPE of two positive norm operators.

Let us now identify the remaining marginal operators in the free boson formulation of the $O(2)$ model with their $n\to2$ limits. Apart from $\veps$ and $\psi_1$, we need another three operators that can only come from operator $\calO_{0,2}$. Besides, two of this operator have to match with some linear combination of $V_{0,4}^\pm$, while the other has to match with the linear combination of $L$ and $V_{1,0}^+$ that is orthogonal to $\veps$. Since $\veps$ is a singlet under the physical $O(2)$, the same must be true for this operator. We will call this operator $S$ for generic $n$ and for $n\to2$ we have $S\to\sqrt{\frac{2}{3}}L+\sqrt{\frac{1}{3}}V_{1,0}^+$.

The operator $\calO_{0,2}$ has multiplicity $M(n)=\frac{1}{4}n(n-1)^2 (n+2)$, which is $2$ for $n=2$.  The scenario we propose is that we can split the multiplicity in $M(n)=f^+(n)+f^-(n)$, where we have $f^+(2)=3$ positive norm operators which do not decouple from the theory at $n=2$ while $f^-(2)=-1$ operators which do decouple, and cancel with $\psi_2$ in the partition function.

The multiplicity $M(n)$ indicates that the operators $\calO_{0,2}$ transform in some reducible representation of $O(n)$. Our rule for decomposing it into irreps of $O(n)$ is that any multiplicity should be a sum of multiplicities of irreps, with non-negative and $n$-independent coefficients \cite{Gorbenko:2018_2,Binder:2019zqc}; this by itself does not identify an unique decomposition in the case of $M(n)$. However, as we just explained at $n=2$, the operators $\calO_{0,2}$ have to give us a singlet $S=\sqrt{\frac{2}{3}}L+\sqrt{\frac{1}{3}}V_{1,0}^+$, and two charge four operator $V_{0,4}^\pm$. Notice that the OPE coefficients match:
\beq
\lambda_{SS\veps}=\lambda_{V_{0,4}^\pm V_{0,4}^\pm\veps}=-\frac{2}{\sqrt{3}}
\eeq
agrees with
\beq
\lim_{n\to 2}\lambda_{\calO_{0,2} \calO_{0,2} \veps} = -\frac{2}{\sqrt{3}}\,.
\eeq

Yet another consistency condition is that  a decomposition of $M(n)$ which satisfies the property of having non-negative, $n$ independent coefficient, which gives a singlet and a charge 4 doublet at $n=2$ exists. Turns out it is also unique:
\beqg
\label{eq:Young}
M(n)=\mathbin{\vcenter{\hbox{\scalebox{1.5}{$\bullet$}}}}+\yng(4)+\left( \yng(2)+\yng(2,2)+\yng(2,1,1)\right) \,,
\eeqg
where we represented $O(n)$ irreps by their Young tableaux ($\mathbin{\vcenter{\hbox{\scalebox{1.5}{$\bullet$}}}}$ is the singlet). The four-indexed irrep symmetric of $O(n)$ gives charge 4 operators in the $n\to 2$ limit and the quantity in parenthesis has multiplicity $-1$ at $n=2$.

Let us summarize the results of this subsection. We found that several nearly marginal operators that exist in the $n\leq2$ theory group together to form five positive norm operators in $n=2$ theory, while remaining operators decouple. Purely from  the OPE data, consistency of $n\to2$ limit, and the requirements of $O(n)$ symmetry we determined that one of the operators $\calO_{0,2}$ is a singlet. These arguments hold both for the critical and low-T fixed points and, importantly, in the latter $\calO_{0,2}$ is relevant.

\subsubsection{Decoupling of non-logarithmic operators}
\label{sec:nonlog}

Let's also consider, as we did in section \ref{sec:mixing}, the spin two operators which have dimension $\frac{5}{2}$ in the $n \to 2 $ limit. We discussed the operator $\Omega$, which is a negative-norm quasi-primary, defined as
\beq
\Omega = (L_{-2}-\alpha L_{-1}^2 )\calO_{0,1}\qquad \text{with norm} \qquad \braket{\Omega \Omega} =- \frac{8}{3m^2}+\ldots\,.
\eeq
As we determined in section \ref{sec:mixing}, this operator mixes with other operators in the theory in the course of the RG flow from the critical to the low-T fixed point which implies that it cannot simply decouple at $n\to2$. Instead
 we will see that a (negative norm) combination of $\Omega$ and $\calO_{2,1}$ decouples, while another remains in the theory. 
Using the $n\to 2$ results, obtained from bootstrap at $n<2$, we get
\beq
\lambda_{\veps \calO_{0,1}\calO_{2,1}}\to \sqrt{\frac{2}{3}} \qquad \lambda_{\veps \calO_{0,1}\Omega}\to-\frac{4}{3\sqrt{3}}\frac{1}{m}
\eeq
we find that the OPE of $\veps$ with $\calO_{0,1}$ contains
\beq
\veps \cdot \calO_{0,1} \supset  \frac{2}{3} \xi_1 \equiv  \frac{2}{3} \left(  \sqrt{\frac{3}{2}} \calO_{2,1}+ \frac{\sqrt{3}}{4}m \Omega \right) \,.
\eeq
It can be seen that $\xi_1$ has norm $\braket{\xi_1 \xi_1}=1$. The orthogonal combination $\xi_2$
\beq
\xi_2 =\frac{1}{\sqrt{2}}\calO_{2,1}+\frac{3 m}{4}\Omega
\eeq
has instead negative norm and drops out of this OPE, and is expected to drop out of any OPE of operators belonging to the unitary closed subalgebra.

In the free boson formulation of $O(2)$, what do these operators match to? We've already seen that in the $n \to 2$ limit $\calO_{0,1}$ matches with $V_{0,2}$. What about the dimension $5/2$, spin 2 operators $\xi_1$ and $\xi_2$? It's easy to identify the vertex operators $V_{1,2}^\pm$, which have the same dimension and spin. There is indeed another operator in the free boson $O(2)$ model, with the same dimension and spin, but it's a bit trickier to identify it. The operator $V_{0,2}$ is a scalar with dimension $\frac{1}{2}$, and it has a null descendant at level two,
\beq
(L_{-2}-L_{-1}^2)V_{0,2}=0\,.
\eeq
Therefore we have another primary operator of spin 2, with the same dimension as $\calO_{2,1}$. This is 
\beq
N^\pm=\frac{1}{\sqrt{3}} \left[(\pd \vphi)^2 V^\pm_{0,2} \mp \frac{1}{\sqrt{2}}\pd^2 \vphi V_{0,2}^\mp \right] \,.
\eeq
This is unit normalized, as can be seen by using formulas \eqref{eq:LVertex}.
Using (see equation \eqref{eq:VVVOpe})
\beq
\lambda_{\veps V^\pm_{0,2}V^\pm_{1,2}}=\frac{1}{\sqrt{3}}\qquad \lambda_{\veps V^\pm_{0,2} N^\pm}=-\frac{1}{3}
\eeq
we find
\beq
\veps \cdot V^\pm_{0,2}\supset \frac{2}{3}\left( \frac{\sqrt{3}}{2}V_{12}^\pm-\frac{1}{2}N^\pm\right) 
\eeq
and we can make the identification $\xi_1 \to  \frac{\sqrt{3}}{2}V_{12}-\frac{1}{2}N$. Notice that also the OPE coefficients match.

What about the orthogonal combination $ \frac{1}{2}V_{12}^\pm+\frac{\sqrt{3}}{2}N^\pm$? Which operator matches it in the $n\to2$ limit? And what is the fate of $\xi_2$? The answer is similar as before.

For $n<2$ we have a spin 2 operator $\calO_{1,2}$, which has dimension $5/2$ in the $n\to2$ limit. Its multiplicity is $n(n+1)^2(n-2)/4$, which goes to zero as $n\to 2$, and that again we cannot decompose it into a unique direct sum of irreps. However, we expect that at $n=2$ we're left with a positive norm operator with multiplicity $2$, which matches with $ \frac{1}{2}V_{12}^\pm+\frac{\sqrt{3}}{2}N^\pm$, and some other operator with multiplicity $-2$ which cancels with $\xi_2$ in the partition function. The requirement that $\calO_{1,2}$ can be decompose into a charge two operator under the physical $O(2)_p$ at $n=2$ does not give a unique decomposition. An example of decomposition that would work is
	\beq
\yng(2)+\left(\yng(3,1)+\yng(2,2)+\yng(1,1,1,1) \right) \,.
	\eeq
	  As a check, we can see that some OPE coefficients match. In particular, as we determined in \ref{sec:mixing},
\beq
\braket{\veps \calO_{0,1} \calO_{1,2}}=\braket{\veps V^\pm_{0,2} \left( \frac{1}{2}V_{12}^\pm+\frac{\sqrt{3}}{2}N^\pm\right) }=0\,.
\eeq
This doesn't mean that $\calO_{1,2}$ decouples from the theory --- there are no reasons for it to --- but we did not compute explicitly its other non-zero OPE coefficients.

To summarize, we observed a very similar mechanism for decoupling of negative norm states both in logarithmic and non-logarithmic sectors. It is tempting to conjecture that this mechanism is quite generic and works for other operators, as well as at other integer values of $n$. Let us mention that one can also consider an extended non-unitary theory which includes the decoupled operators. Generically, this theory will be logarithmic \cite{Cardy:2013rqg}. Somewhat surprisingly, torus partition functions of the two theories are identical.

\section{Logarithms in the two-dimensional Potts model}
\label{PottsLogs}
We gave several examples of logarithmic operator in the two-dimensional $O(n)$ model. A natural question to ask is how specific to this model is the existence of logarithmic operators. Another interesting model, which shares many similarities with the $O(n)$ one, is the critical Potts model in two dimension, a model with $S_Q$ symmetry, where $Q$ is a parameter that we can change continuously. It has been studied intensively for decades, see for example \cite{Nienhuis1987,diFrancesco:1987qf,baxter_book,Jacobsen2012}, and recently, by us \cite{Gorbenko:2018_2}. It was shown that, analogously to the $O(n)$ model, 2d Potts is logarithmic for a discrete infinite set of $Q$ \cite{Couvreur:2017inl}, and it was suspected to be logarithmic for every value of $Q$ \cite{Jacobsen:2018pti}, however, to the best of our knowledge, this was not shown explicitly.\footnote{Logarithmic operators were observed in a non rigorous manner, similar in spirit to our \ref{sec:log_limits}, for generic values of $Q$ in \cite{Viti3} by using Coulomb gases techniques. It is also worth mentioning that logarithmic operators were also found explicitly in the case of $Q=1$ and $Q=2$ with boundaries \cite{Viti1,Viti2}.}
Now that we have more familiarity with logarithmic CFTs, we briefly look at the model again and prove that the two-dimensional critical Potts is indeed a logCFT for generic $Q$.

We briefly mention our conventions, which can be found in Table 1 of \cite{Gorbenko:2018_2}. We look at the Potts model for $0 < Q \le 4$, for which the CFT exists at real values of the coupling \cite{Baxter:2000ez}.\footnote{For $Q > 4$, see \cite{Gorbenko:2018_1,Gorbenko:2018_2}.} The central charge of the model is
\beq
c_P=1-\frac{6}{m_P(m_P+1)} \qquad \text{with} \qquad m_P=\frac{2 \pi }{\cos ^{-1}\left(\frac{Q}{2}-1\right)}-1\,.
\eeq

The energy operator has a null descendant at level 2, $\veps_P=\phi_{2,1}$, as can be seen by expanding the torus partition function \cite{diFrancesco:1987qf}, and noticing that the partition function contains the term $q^{h_{2,1}}\bq^{h_{2,1}+2}$ with multiplicity one.\footnote{In this section $h$ is still defined in the usual way \eqref{eq:hKac}, but with $m_P$ instead of $m$.}
Therefore, its correlation functions satisfy a differential equation similar to \eqref{eq:diffeq4eps}, with the difference that it will be second order this time.

Then let's consider the scalar operator $\calO_{0,1}$, following the notation of \cite{diFrancesco:1987qf}, which has dimension $(h_{0,2},h_{0,2})$. This operator transforms in the two indexed symmetric representation of $S_Q$ and has multiplicity $\frac{Q(Q-3)}{2}$. By imposing crossing on the correlator $\braket{\calO_{0,1}\calO_{0,1}\veps\veps}$, we find that logarithmic operators are exchanged in the $\calO_{0,1} \cdot \veps$ OPE. To be precise, we find a level 2 staggered module for the operator $\calO_{2,1}$, which has dimension $(h_{1,2},h_{-1,2})$, and has a zero norm descendant at level 2. Expanding the partition function we can check that the number of operators is what is to be expected from the logarithmic structure. 

\begin{figure}
	\centering
	\includegraphics[scale=.7]{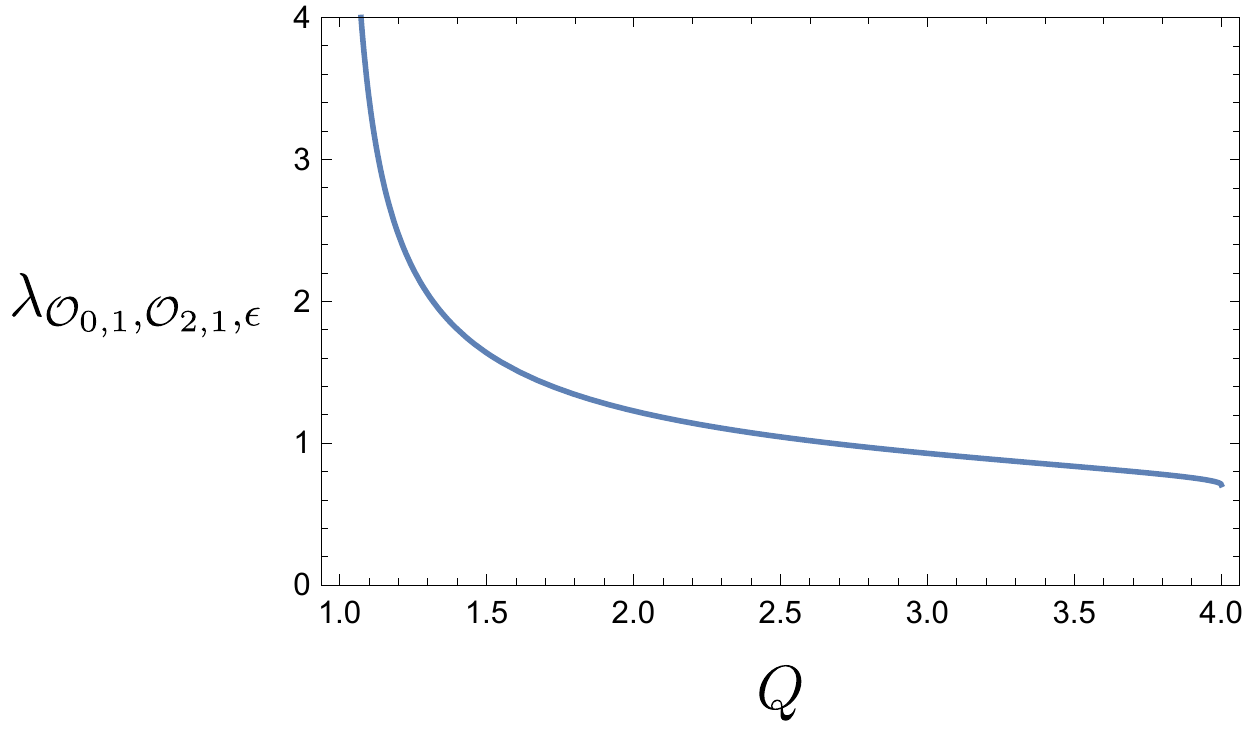} \hspace{4em}
	\caption{The OPE $\calO_{0,1} \veps$ contains the field $ \calO_{2,1}$ which is part of a level two staggered module. We plot here the OPE coefficient. In a small $z,\bz$ expansion, the $\braket{\veps(0)\calO_{0,1}(z,\bz)\calO_{0,1}(1)\veps(\infty)}$ four point function will have logarithms of $z \bz$ with prefactors proportional to $\lambda_{\calO_{0,1}\calO_{2,1}\veps}^2$.}
	\label{fig:Potts_logs}
\end{figure}

We also mention that we expect the correlation function $\braket{\calO_{0,k}\calO_{0,k}\veps\veps}$, with $k$ half integer, $k\ge1$, to exchange logarithmic operators in the t-channel, although we have not checked explicitly what the implications of crossing symmetry are. Operators $\calO_{2,k}$, with dimension $(h_{1,2k},h_{1,-2k})$, might be exchanged in this channel and we expect them to be part of a level $2k$ staggered module. 
We show the value of the OPE coefficient of an operator in a logarithmic multiplet  in the range $1\le Q \le4$ in figure \ref{fig:Potts_logs}.

After this preprint was finished, the paper \cite{He:2020rfk} appeared, in which it's stated that operators whose dimension is part of the Kac table, and that are not $O(n)$ singlets, are expected to be in general logarithmic. This agrees with our identification of $\calO_{2,k}$ as potential candidates and with our explicit computation in the case $k=1$.

\section{Conclusions and open questions}
\label{sec:conclusions}
Let us briefly summarize our findings and present a list of problems related to $O(n)$ models that remain unsolved. We have studied the critical  and low-temperature $O(n)$ models in two dimensions for continuous $n$ with $n<2$, and found that they are logarithmic CFTs. Using the knowledge of the torus partition function, as well as properties of the $O(n)$ categorical symmetry, we've determined that the energy operator is degenerate and has null descendants. Hence correlation functions containing this operator satisfy third order BPZ differential equations. This allowed us to fix some of the OPE coefficients of the theory, as well as to identify several logarithmic multiplets. The most notable example is the current $J^\mu$, which is part of a level 1 staggered multiplet. Using the conformal data, we've studied the RG flow from the critical to the low-T fixed point explicitly to first order in conformal perturbation theory, finding perfect agreement. Finally, we've studied how logarithmic and negative norm operators drop out of the theory in the $n \to 2$ limit, giving rise to a unitary subsector, which corresponds to the BKT critical theory described by a compact boson. By demanding the smoothness of the $n\to2$ limit we determined additional CFT data at $n<2$, and in particular $O(n)$ quantum numbers of some of the low-lying operators. Among these operators, we found an $O(n)$ singlet, known to correspond to a possibility of loop crossings in the related $O(n)$ invariant loop models. 

We now discuss several puzzles and possible future directions that we find interesting.
One natural question to ask at this point is which properties of the 2d $O(n)$ model carry on to higher dimensions. As we already mentioned, definition of corresponding loop models for any $n$ \cite{3dON} is readily available in 3d for any $n$ (see also \cite{Nahum_2013} for other examples of 3d loop models); generalizations to higher dimensions are also possible. For some integer values of $n$ the model is known to be logarithmic \cite{Cardy:2013rqg}, while it is an open question if the same holds for generic $n$. We don't have anything to say about this at this point because we do not have a proper understanding of the underlying reason for appearance of the logs even in 2d. What we found is that  the current operator is logarithmic and hence generic correlators of charged operators will also be. Most operators in the loop formulation of the $O(n)$ model have a geometric meaning \cite{Nienhuis1987,cardy1994geometrical}. It is thus tempting to develop a geometric intuition lying behind these logarithms. One thing we would like to point out is that staggered modules seem to be less likely to appear in higher dimensions. In a staggered module, we need to have a zero norm descendant; in the case of the current module, for example, this operator was $B$. This requirement is highly non-trivial, and in two dimensions it translates to having an operator whose weight is part of the Kac table, see for example the operators $\calO_{2,k/2}$.

In higher dimensions, we have less operators that have a zero norm descendant, so it looks harder to have a staggered module. However, this does not mean that it's impossible: for example, we know of an example of staggered module in $d=6$ \cite{Brust:2016gjy}. For what concerns the currents in higher dimensions, we don't expect them to be part of a logarithmic multiplet: we only have one operator $J_\mu$ rather than $J$ and $\bJ$, and we expect it to be conserved, $\pd_{\mu} J^\mu=0$, in the same way as a linear combination of $\pd \bJ$ and $\bpd J$ is conserved in two dimension. On the other hand, the structure of ordinary, non-staggered, logarithmic modules do not seem to depend strongly on dimension.

Usually the presence of logarithmic terms in the theory is related to tuning of some parameter. Since we found logs for a generic $n$, it could be that such parameter in our case is dimensionality of space $d$. If this is the case, this brings another natural questions: whether continuous-n $O(n)$ models allow a natural extension to non-integer dimensions. In the continuum limit such extension can be provided by the epsilon expansion around $d=4$, or $d=2$, combined with analytic continuation in $n$. We note that the status of such continuation is different from what we have in integer $d$, where the microscopic theory is well-defined.
 Independently of this, it would also be interesting to see if for any $n$ and $d$ other than 2 $O(n)$ models have critical points analogous to the low-T phase, thus completing the $d-n$ phase diagram conjectured long time ago in \cite{PhysRevLett.45.499}.

We also note that theories either at non-integer $n$ \cite{Maldacena:2011jn,Binder:2019zqc} or at non-integer d \cite{parisi2002spaces,Hogervorst:2015akt} are necessarily non-unitary and contain negative-norm states. It is thus interesting to explore whether the mechanism of decoupling of these states at integer $n$ and $d$ is always similar to the one studied by us in section~\ref{sec:nonlog} for non-logarithmic operators.

A question which we dodged in the present work concerns the stability of logarithmic multiplets under perturbations. If we perturb a logCFT by a weakly relevant perturbation so that we flow to an IR CFT, does the logarithmic structure of the theory survive? Intuitively, we expect non-logarithmic degenerate operators to acquire different anomalous dimension under RG flows simply because of eigenvalue repulsion. However, we've seen by direct inspection that the current (as well as the $\calO_{2,k/2}$ operators) is part of a logarithmic multiplet in both the critical fixed point and in the low temperature fixed point, so the logarithmic structure survives the RG flow. To answer this question one needs to set up conformal perturbation theory even for logarithmic multiplets. We only made one step in this direction by writing down the structure of logarithmic three-point functions for some of our operators.

In this paper we only used the most basic constraints of the categorical $O(n)$ symmetry \cite{Binder:2019zqc}. Naturally, one could study exploit many more of such constraints if one studied correlation functions of operators transforming in more non-trivial representations of $O(n)$. This would require bootstrapping correlators without an insertion of $\veps$. The same questions apply to the  closely related $Q$-states Potts model which has categorical $S_Q$ symmetry. Some steps in this direction were made in \cite{Picco:2016ilr} (see also \cite{Jacobsen:2018pti}).

Yet another avenue to pursue is to couple these models to 2-dimensional gravity \cite{Kostov:1988fy,Eynard:1995nv,Eynard:1995zv}. It is interesting to see if any of the novel features of the model that we discovered have any implications in the gravitational case.

Finally, let us mention two open problems on which we were able to make at least some progress.

\subsection{``Dangerously irrelevant'' singlet operator}
In this section we summarize our findings related to the ``dangerously irrelevant'' $O(n)$ singlet operator present in our theory. This operator has dimension of $\calO_{0,2}$, which for $n$ close to 2 reads $2\pm\frac{2}{\pi}\sqrt{2-n}$ with $+$ sign corresponding to the critical (UV) fixed point and $-$ to the low-T (IR) fixed point. That is, in the process of the RG flow this operator switches from being irrelevant to relevant. The total multiplicity of the operator $\calO_{0,2}$ in the partition function is $M(n)=\frac{1}{4}n(n-1)^2 (n+2)$. In section~\ref{sec:n->2}, by matching all operators which become marginal in the $n\to2$ limit to the known spectrum of the $O(2)$ model, we determined that one of the operators with dimension of $\calO_{0,2}$ must be a singlet of the categorical $O(n)$ symmetry, and the unique decomposition of $M(n)$ into irreducible representations is given by \ref{eq:Young}. We stress that the only assumptions that went into proving this statement were $O(n)$ symmetry, crossing relations and continuity of correlation functions in the $n\to2$ limit. We also calculated the following OPE coefficient of this singlet operator, which we called $S$:
\beq
\lambda_{SS\veps}=\frac{-2}{\sqrt{3}}\,,\quad \lambda_{SSS}=\frac{-2\sqrt{2}}{\sqrt{3}}\,, \quad  \lambda_{S\veps\veps}=0\,,
\eeq
up to $O(\sqrt{2-n})$ corrections.

Let us now explain why this operator presents a puzzle. From all that is known, it is singlet under all global internal symmetries of the theory, and usually such operators get generated in the course of the RG flow. Given that it is relevant in the IR fixed point, this would imply that it is unstable and one would need to tune some combination of operators $S$ and $\veps$ near the UV fixed point in order to reach the IR. Nevertheless, this is not what happens: the RG flow triggered just by $\veps$ (with an appropriate sign of the deformation) reaches the IR fixed point. At the leading order in conformal perturbation theory, this follows from vanishing $\lambda_{S\veps\veps}$, however, this appears to be true non-perturbatively for finite $2-n$. In fact, genericity of the low-T fixed points holds not just for the RG flows from the critical fixed point, but also for many classes of $O(n)$ loop models, both on honeycomb and square lattices \cite{Blote1989,Blote_square,PhysRevE.83.021115}, and is believed to hold as long as the loop model does not allow loop-crossings. It is tempting to attribute this effect to some hidden symmetry, namely that the operator $S$ corresponds to the loop-crossing term in the lattice theory and that both are protected by some hidden symmetry of the model. 

Indeed, there are known examples of loop models with enhanced symmetry (often from $O(n)$ to $U(n)$) where the loop crossing terms breaks the enhanced symmetry \cite{PhysRevB.87.184204,Vernier:2016erq,Nahum:2015gjk}. It was also shown in \cite{Read:2001pz} that the low-T phase of the $O(n)$ model for some integer $n$ can be represented as a $U(n)$-symmetric model with an addition of some topological defects. This lead reference \cite{Jacobsen:2002wu} to conjecture that for these integer $n$'s $U(n)$ invariance also guarantees the genericity of the low-T phase of the $O(n)$ model. For this to be the case, of course, $U(n)$ symmetry must be the global symmetry not just in the low-T phase, but also in the critical phase and of any lattice formulation of a non-intersecting $O(n)$ model. It is an interesting proposal for selected integer values of $n$, but, while a possibility, we would find it surprising if it was the case for generic $n$ for the following reason.
 We cannot simply fit any additional continuous symmetry into the model because it would correspond to additional conserved currents, while we have exactly $n(n-1)/2$ of those in the theory. This means that the extra currents must be completely hidden in the torus partition function because of an exact cancellation between different topologically twisted sectors for any $n$. The same cancellations must be present for other operators that form complete representations of $O(n)$ but not of $U(n)$. 
 
  If, instead, there was some additional discrete symmetry, $S$ would have to transform in a one-dimensional representation (e.g. multiplication by a number), but this contradicts having a non-zero OPE coefficient $ \lambda_{SSS}$. We thus conclude that whatever is protecting operator $S$ cannot be a usual global symmetry, even as a categorical symmetry in a sense of \cite{Binder:2019zqc}.\footnote{There is an unlikely to us possibility that the $O(n)$ symmetry of the model is somehow broken down to a smaller subgroup, for example to $U(n/2)$. This could allow $S$ to transform non-trivially under this smaller symmetry and thus be protected. We could not completely exclude this possibility, but we did find multiple consistency checks of the presence of the $O(n)$ symmetry in the model.}

One possible resolution lies in a fact that $S$ is protected by a space-time symmetry related to integrability. Indeed, operator $\veps$ commutes with a subset of Virasoro charges which guarantees integrability of the RG flow triggered by it \cite{Zamolodchikov:1987jf}. Operator $S$ indeed breaks those charges. This, however, is not enough to explain why multiple non-intersecting loop models also flow to the low-T fixed points. Some of those models can indeed be transformed to integrable vertex models, but it is not clear whether all non-intersecting loop models flowing to the low-T fixed point are integrable and whether such integrability is sufficient to explain the genericity of the low-T fixed point. At the same time, if a lattice model does allow loop crossing, the partition function in terms of loops has extra terms \cite{Chayes} and it is not known how to map it exactly to a vertex model. It would be interesting to investigate this further.

At this moment, the puzzle remains unresolved. Let us bring up a couple of examples in the literature which appear to have some resemblance to our situation. A recent interesting paper \cite{Chang:2018iay} studied several two-dimensional RG flows in which relevant operators are protected by so-called non-invertible topological defect lines. While invertible lines correspond to usual global symmetries, non-invertible lines do not. Nevertheless, they are preserved by RG and prevent local operators that do not commute with them from being generated. Similar phenomena are present in 2d adjoint QCD with massless quarks, where one of the two singlet four-fermion operators is also protected by such lines \cite{Zohar}. It is an exciting possibility if, in addition to the categorical $O(n)$ symmetry, lattice loop models without crossings and the CFTs we studied in this work also contain some similar objects that protect operator $S$.

\label{sec:Dangerous}

\subsection{The $n >2 $ region, periodic S-matrices and complex CFTs}
\label{sec:S-matrix}
In this work we focused on the $n \lesssim 2$ regime, but
part of our motivation to study two-dimensional $O(n)$ models was that for $n>2$ they provide an example of walking RG flows. Here we can think of the critical and the low-T fixed points as living at complex values of the coupling, and being described by complex CFTs \cite{Gorbenko:2018_1,Gorbenko:2018_2}. We reserve detailed investigation of this to future work, but as an invitation to the reader we show a potential relation with the periodic $O(n)$ symmetric S-matrix, a solution to Yang Baxter equation originally written down in \cite{Hortacsu:1979pu} and recently rediscovered in the S-matrix bootstrap program \cite{Cordova:2018uop,Cordova:2019lot}.

\begin{figure}
	\centering
	\begin{subfigure}{.5\textwidth}
		\centering
		\includegraphics[width=.8\linewidth]{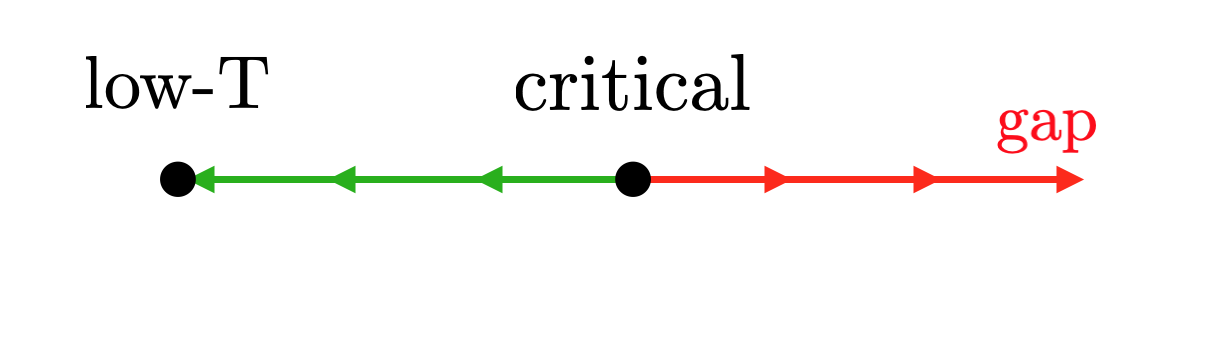}
		\caption{$n<2$}
	\end{subfigure}%
	\begin{subfigure}{.4\textwidth}
		\centering
		\includegraphics[width=.8\linewidth]{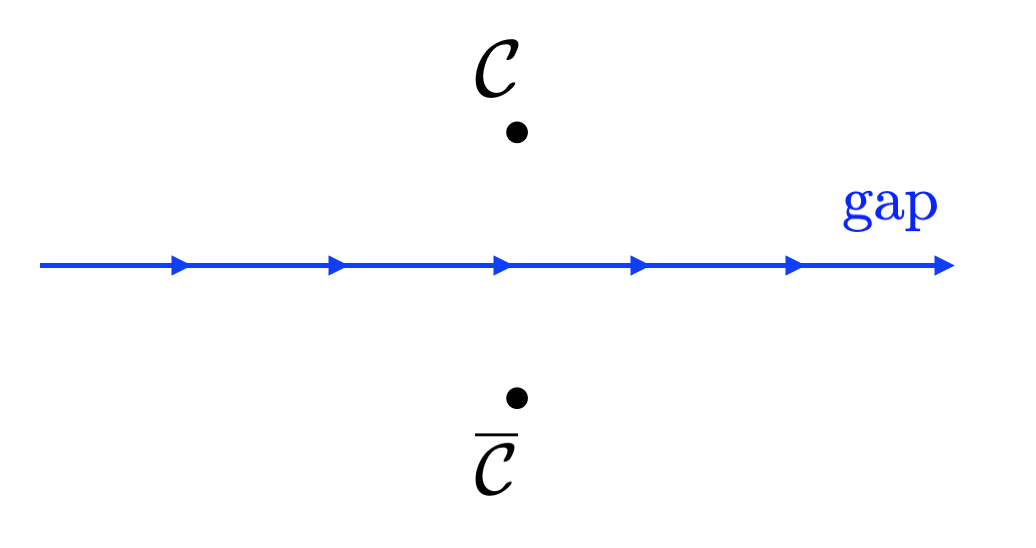}
		\caption{$n>2$}
	\end{subfigure}
	\caption{RG flows triggered by deformation of $O(n)$ CFTs by $\veps$. Green -- positive coupling, flow studied in section \ref{sec:flow}; red -- negative coupling, flow studied in \cite{Zamolodchikov:1990dg}; blue -- complex coupling deformation of an $n>2$ complex CFT.  }
	\label{fig:flows}
\end{figure}

Let us first review the known S-matrix which describes the massive integrable RG flow from the critical $n\less2$ fixed point deformed by $\veps$, albeit with an opposite sign of the coupling than we considered above.  This S-matrix was  defined in \cite{Zamolodchikov:1990dg} and it has the striking feature that forward scattering is not allowed in it. Particles can only ``reflect'' or ``annihilate'' with each other. The non-crossing of particle worldlines reminds of the mutually avoidance of the loops in the $O(n)$ model. The S-matrix studied in \cite{Hortacsu:1979pu,Cordova:2018uop,Cordova:2019lot} is defined for $n>2$ and has the same feature, which suggests that it may also have something to do with the loop models, and hence with our complex CFTs. In the singlet channel it takes the following form\footnote{Compared to eq.(35) of \cite{Cordova:2018uop}, we consider an overall minus sign in front of the S-matrix in order to have this matching. This minus sign is innocuous but it does change physical properties of the theory. The theory with the sign of \cite{Cordova:2018uop} also appears to walk, but near a different value of $c$.}
\beq
\label{Ssing}
	S_{sing}=-e^{-i k \theta} \prod_{l=1}^\infty \frac{\sinh[k(i \theta -2 l \pi)]\sinh[k(i \theta +(2 l+1) \pi)]}{\sinh[k(i \theta +2 l \pi)]	\sinh[k(i \theta -(2 l+1) \pi)]}\,,\quad k=\frac{\text{arccosh}\frac{n}{2}}{\pi}\,.
\eeq
 For $n=2$ this S-matrices coincides with the $n \to 2$ limit of the $n\leq2$ one, however, they are not related by a simple analytic continuation. Indeed this is not to be expected: $n\leq2$ S-matrices possess a UV fixed point, while we do not expect this to be the case for $n>2$ theories. In fact, $n>2$ S-matrices have a very unusual UV behavior: 
they are periodic under real shifts of the rapidity $\theta \to \theta +2 \pi^2/\text{arccosh}\frac{n}{2}$, which at high energies implies periodicity in the logarithm of the energy scale. We thus conjecture that periodic S-matrices with $n\gtrsim2$ ``walk'' for some long RG time between the complex $O(n)$ fixed points, but never reach them. In this walking regime the physics of the S-matrix theory can be obtained from conformal perturbation theory around one of the complex CFTs, deformed by operator $\veps$ with a complex coupling, see figure \ref{fig:flows}. For integer values of $n$ we expect the deformation to restore unitarity, while for non-integer $n$ deformed theory will be real but non-unitary. 

\begin{figure}
	\centering
	\begin{subfigure}{.5\textwidth}
		\centering
		\includegraphics[width=.8\linewidth]{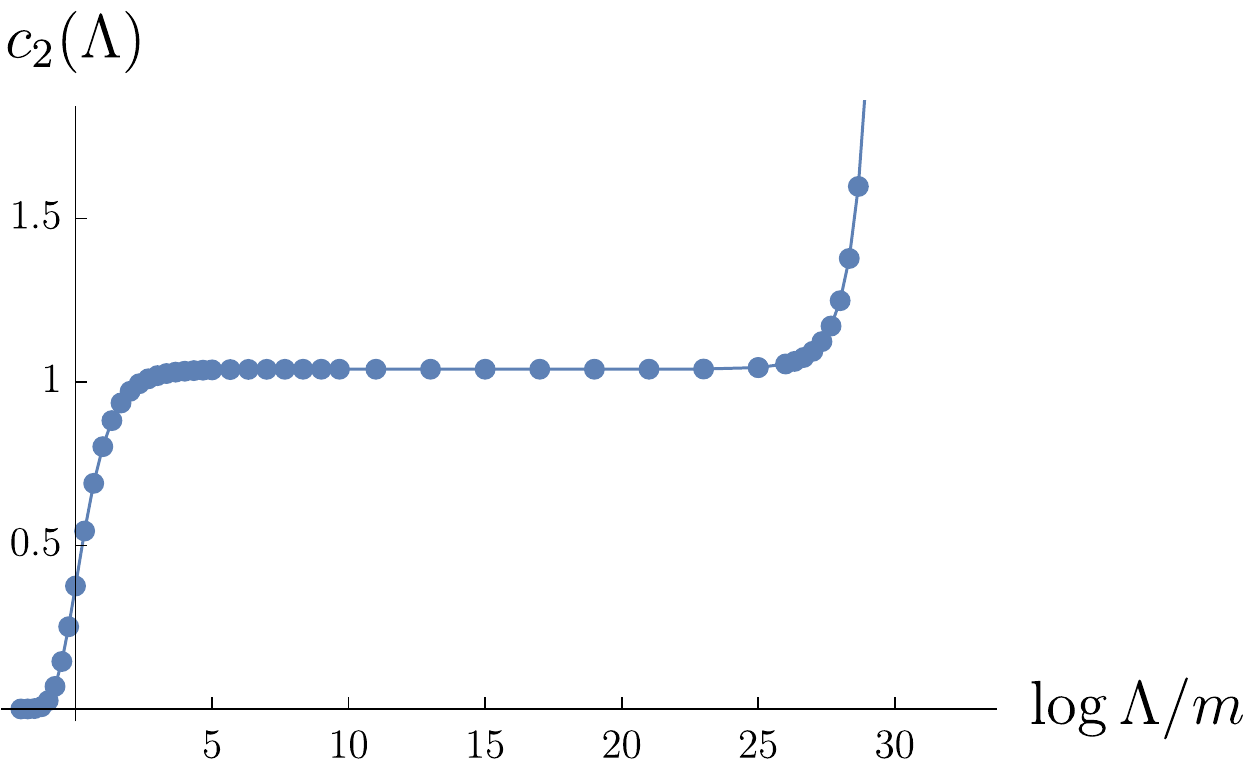}
		\caption{$n=2.1$}
	\end{subfigure}%
	\begin{subfigure}{.5\textwidth}
		\centering
		\includegraphics[width=.8\linewidth]{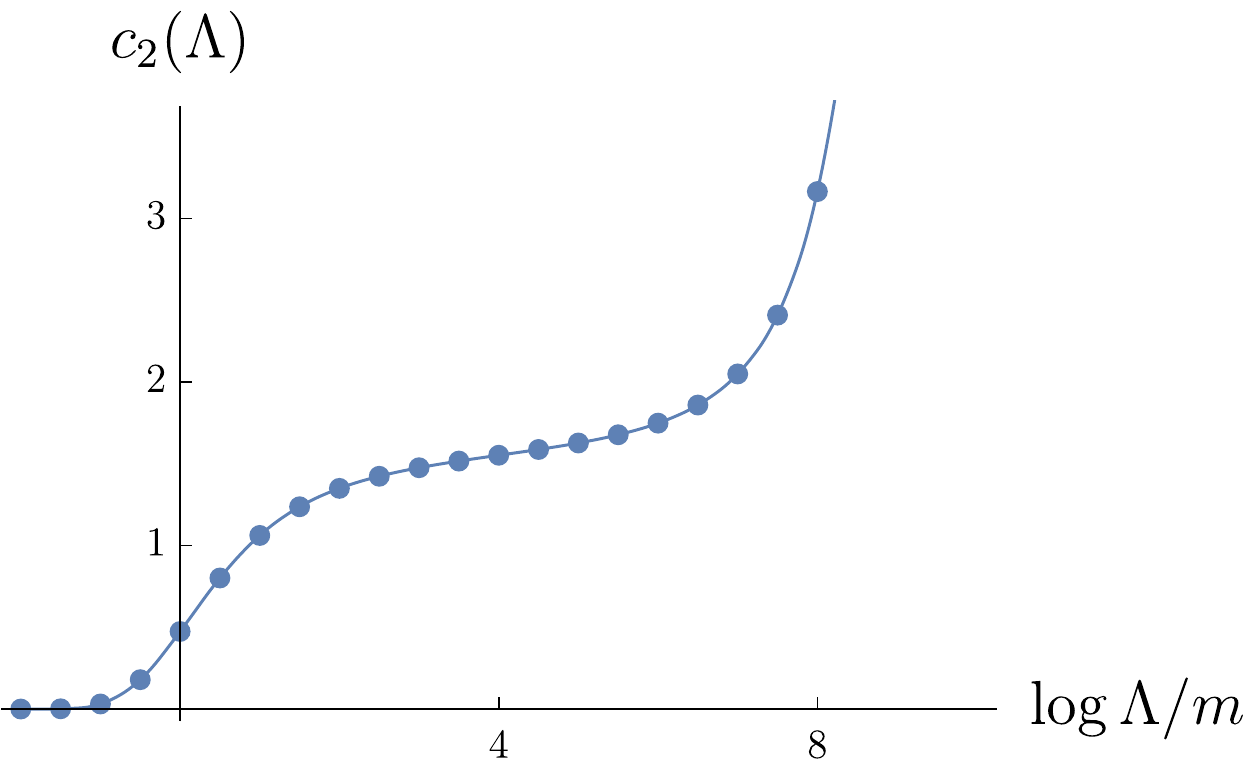}
		\caption{$n=3$}
	\end{subfigure}
	\caption{The value of $c$ along the RG flow. The mass $m$ is the mass of the particles in the fundamental of $O(n)$.}
	\label{fig:c_walk}
\end{figure}

\begin{figure}
	\centering
\includegraphics[scale=.5]{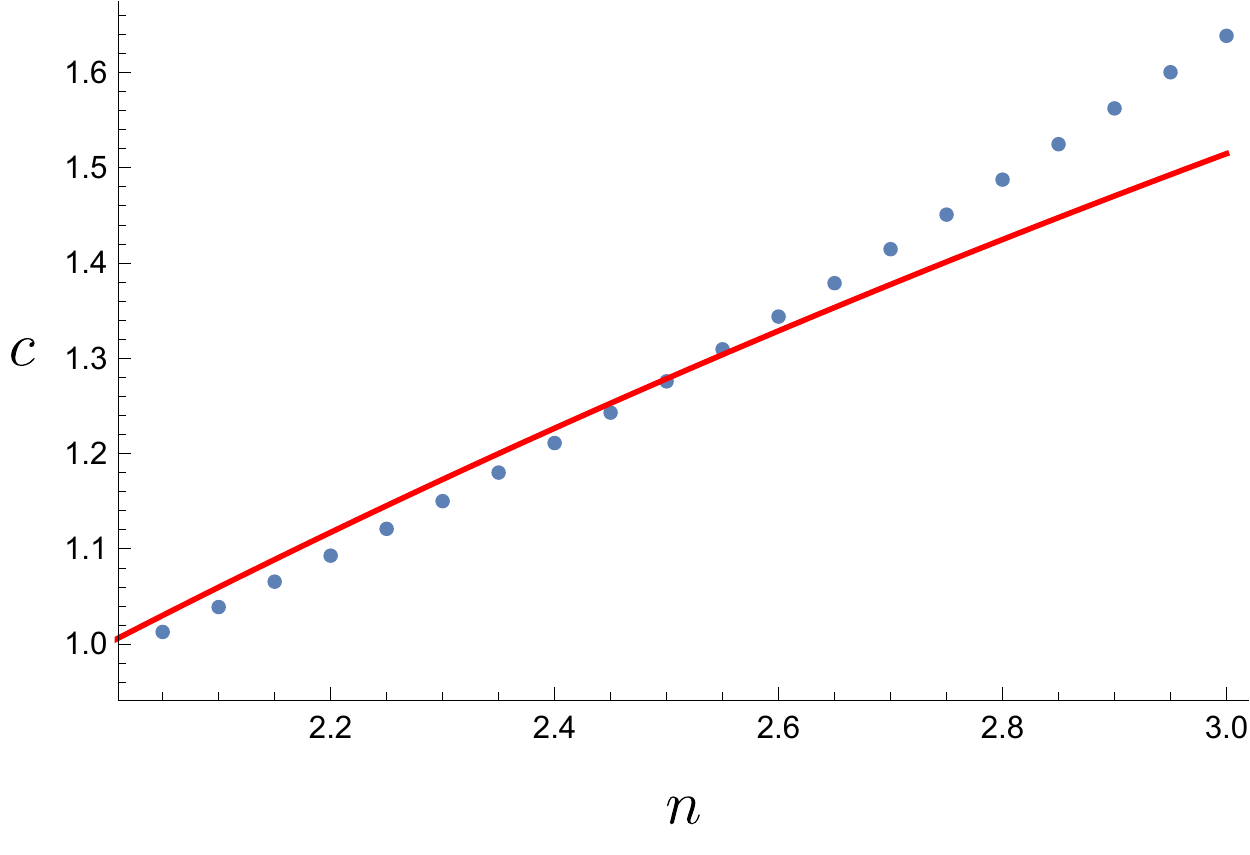}
\caption{The value of $c(\Lambda)$ in the middle of the plateau (blue dots) compared with the real part of $c$ at the complex fixed points (red line). The discrepancy is small for $n$ close to 2, where the complex CFTs are close to the real axis. As $n$ increases, the complex fixed points move further away from the real axis and the c-function of the walking theory starts to differ from the real part of the central charge of the complex CFTs. At this point one needs to include corrections from conformal perturbation theory.}
\label{fig:c_re}
\end{figure}

To test this proposal, we use the form-factor approach \cite{Karowski:1978vz,10010364330} to compute the two point function of $\Theta$ along the flow and infer from it the c-function of the theory \cite{Zamolodchikov:1986gt,PhysRevLett.60.2709}:\footnote{See for example \cite{Karateev:2019ymz} for a propaedeutic explanation of the needed techniques.}
\beq
\label{clambda}
c(\Lambda)=6\pi^2 \int^\infty_{\frac{1}{\Lambda}}d x \braket{\Theta(x)\Theta(0)} =c_2(\Lambda)+\ldots\,,
\eeq
where $\Theta$ is the trace of the stress energy tensor, and the dots represent higher particle contributions which we are neglecting. This approach was used in \cite{Cardy_1993} to compute the same object in the $n<2$ theory and was shown to produce very precise results. We report the value of $c(\Lambda)$ computed numerically for two values of $n$ in figure \ref{fig:c_walk}.
For $n \gtrsim 2$, we find this quantity to be approximately constant in some range $\Lambda_{IR}<\Lambda<\Lambda_{UV}$. This is what is expected from walking behavior, and we also checked numerically that $\log \Lambda_{UV}/\Lambda_{IR} \sim \frac{\pi^2}{\sqrt{n-2}}$.  By analytically continuing the dimension of operators and OPE coefficients in the region $n>2$  and using conformal perturbation theory at first order we can compute 
\beq
\log \Lambda_{UV}/\Lambda_{IR} =\frac{2 \pi}{|\text{Im}\Delta_\veps|}= \frac{\pi^2}{2\sqrt{n-2}}+O(1)\,.
\eeq
We see that the power of $\sqrt{n-2}$ matches, however, the coefficient differs roughly by a factor of two. The height of the plateau, in turn, should be compared to the real part of the central charge of the complex CFTs, given by the analytic continuation of \ref{eq:c}.  We show the comparison in figure \ref{fig:c_re} and observe a relatively good match. 
The most obvious explanation of the mismatch in the range of walking comes from neglecting the higher particle form factors. Usually they play an increasingly important role in the UV and can easily change the range of walking; we don't expect, however, them to influence the value of $c(\Lambda)$ for which we have walking, since the plateau extends to the IR, where the two particle form factor should be a good description of the theory. For example at $n=2$, where the UV fixed point still exists and $\Lambda$ in \eqref{clambda} can be taken to infinity, two-particle form-factors give $c_{UV}\approx0.98$ and, since all higher-particle contributions must be positive, none of them can be larger than $0.02$. Above the walking region the central charge blows up, and we don't see any UV fixed point. In fact, once on the plateau, the S-matrix for the physical values of the rapidity takes a relatively simple form:
\beq
S_{sing}\simeq -e^{- i k \theta}\,.
\eeq
It remains unknown whether such UV behavior leads to a pathology, is specific to the model at hand, or corresponds to a novel class of non-Wilsonian UV-completions in the spirit of asymptotic fragility \cite{Dubovsky:2013ira}.
It would be interesting to understand better the physics of this peculiar S-matrix, as well as test further our conjectured relation to the complex $O(n)$ CFTs by computing correlation functions of some  other operators in the walking regime. 

There is, of course, a much better known $O(n)$ invariant integrable S-matrix -- the $O(n)$ sigma model \cite{ZAMOLODCHIKOV1979253}. It generically contains forward scattering and, consequently, cannot correspond to a non-intersecting loop model. Neither this S-matrix exhibits any visible signs of walking. Nevertheless, for $n\to2$ it coincides with the periodic one, and thus it should also inhabit the vicinity of the theory space that we study.

 \section*{Acknowledgments}
 We thank Jesper Jacobsen, Denis Karateev, Shota Komatsu, Petr Kravchuk, Slava Rychkov, and Hubert Saleur for discussions and John Cardy and Slava Rychkov for comments on the draft. 
 The work of BZ was supported in part by the US NSF under Grant No. PHY-1914860. 
VG is partially supported by the Simons Foundation Origins of the Universe program (Modern Inflationary
Cosmology collaboration).
 
\appendix

\section{Explicit formula of some OPE coefficients for generic $n$}
\label{app:minimalOPE}
We report here some of the OPE coefficients we found for generic $n$. We found these OPE coefficients for several values of $m$ and noticed that some of these formulas are the same as in diagonal minimal models \cite{Dotsenko:1984ad} while others have minor differences which we were able to guess. The function $C$ is defined in equation (A.5) of \cite{Poghossian:2013fda} (the order of indices is important)
\beqg
\lambda_{\veps\veps\veps}=C_{(1,3),(1,3),(1,3)}\\
\lambda_{\sigma \sigma \veps}=-C_{(1,3),(\frac{m-1}{2},\frac{m+1}{2}),(\frac{m-1}{2},\frac{m+1}{2})}\\
\lambda_{\calO_{0,k/2} \calO_{0,k/2} \sigma \veps}=-\lim_{j\to k}C_{(1,3),(j(m-1),j(m+1)),(j(m-1),j(m+1))}\\
\lambda_{J \bJ\veps}=i C_{(1,3),(1,-1),(1,1)}\\
\lambda_{JJ\veps}=\frac{h_\veps-1}{h_\veps} \lambda_{J \bJ\veps}\,.
\eeqg
\section{Some $n=2$ formulas} \label{app:n2explicit}
In this appendix, we list some explicit four point functions for the $n=2$ theory. The correlator of four energy operators and of two currents and two energy operators are already given in section \ref{sec:crossing}. 
The correlator of two spins and two energy operators at $n=2$ is 
\beq
\braket{\sigma(0)\sigma(z,\bz)\veps(1)\veps(\infty)}=\frac{1}{(z \bz)^{1/8}} \left(\calV_{\id}(z) \calV_{\id}(\bz)+ \lambda_{\veps \veps \veps} \lambda_{ \sigma \sigma \veps} \calV_{\veps}(z) \calV_{\veps}(\bz)+ \lambda_{\veps \veps \veps'} \lambda_{ \sigma \sigma \veps'} \calV_{\veps'}(z) \calV_{\veps'}(\bz) \right) 
\eeq
where
\beqg
\calV_{\id}(z)=\frac{1}{1-z}\left[\frac{1}{3}(2-z)\sqrt{1-z}+\frac{1}{24}(z^2-8z+8) \right] \\
\calV_{\veps}(z)=\frac{z}{\sqrt{1-z}}\\
\calV_{\veps'}(z)=\frac{1}{1-z}\left[64(z-2)\sqrt{1-z}+16 (z^2-8z+8) \right]
\eeqg
and crossing fixes the OPE coefficients to be
\beq
\lambda_{ \veps \veps \veps}\lambda_{ \sigma \sigma \veps}=-\frac{1}{6}\qquad  \lambda_{\veps \veps \veps'} \lambda_{ \sigma \sigma \veps'} =\frac{1}{73728}\,.
\eeq

The correlator of two currents of different helicity and two energy operators acquires a rather simpler form,
\beq
\braket{J(0) \bJ(z,\bz) \veps(1)\veps(\infty)}=\frac{c}{1-\bz}
\eeq
Just by considering crossing we cannot fix the overall normalization $c$, but this can be fixed to be $-4/3$ by an explicit computation for the $O(2)$ model.
\beq
\eeq

\subsection{Free boson OPE coefficients} \label{sec:O2}

We start with the action
\beq
S=\frac{1}{8\pi}\int d^2 x (\partial_\mu \phi)^2\,,
\eeq
which gives the correlator
\beq
\braket{\phi(z,\bz) \phi(0)}=-\log|x|^2\,.
\eeq

The three point function of a current and two vertex operators is
\beq
\langle \pd \vphi(x_1)V_{n,m}^+(x_2) V_{n,m}^-(x_3) \rangle=p \frac{x_{23}}{x_{13}x_{12}}\frac{1}{|x_{23}|^{2p^2}}\,,
\eeq
where $p$ is related to $n,m$ by \eqref{eq:pnm}.

Acting with $L_{-1}^2$ and $L_{-2}$ on a vertex operator gives
\beqg \label{eq:LVertex}
L_{-2}V_{n,m}^\pm = -\frac{1}{2}\left( \pd \vphi \right)^2V_{n,m}^\pm \mp p \pd^2 \vphi V_{n,m}^\mp \\
L_{-1}^2V_{n,m}^\pm = -p^2\left( \pd \vphi \right)^2V_{n,m}^\pm \mp p \pd^2 \vphi V_{n,m}^\mp\,,
\eeqg
which means that we can find correlation functions of $(\pd \vphi)^2 V_{n,m}$ or $\pd^2 \vphi V_{n,m}$ by acting with some differential operator on correlations functions of $V_{n,m}$.

The OPE coefficient of three vertex operators is
\beqg
\lambda_{V_{n,m}^+V_{n',m'}^+V_{n\pm n',m\pm m'}^+}=\frac{1}{\sqrt{2}}\\
\lambda_{V_{n,m}^-V_{n',m'}^-V_{n\pm n',m\pm m'}^+}=\mp \frac{1}{\sqrt{2}} \label{eq:VVVOpe}
\eeqg
for $n \pm n'\neq0$ and $m \pm m'\neq0$.

\section{Logarithmic three point functions} \label{sec:log3pfs}
In this section we look at three point functions with up to two insertion of a logarithmic field.
Three point functions have logarithms only when $A$ is inserted. By looking at the ordinary ones (no insertion of $A$), we find the following relations between OPE coefficients simply by using the fact that $B$ is a descendant of $J$, $B =\frac{1}{2s}\pd \bJ$,
\beqg
\lambda_{B J \veps}=\frac{h_\veps}{2s}\lambda _{J J \veps} \\
\lambda_{B B \veps}=\frac{(1-h_\veps)h_\veps}{4s^2}\lambda _{J J \veps}\\
\lambda_{J \bJ \veps}=\frac{h_\veps}{h_\veps-1}\lambda_{J J \veps}\label{eq:BOPE}
\eeqg
where the last relation follows from the fact that $B$ can be obtained both as a descendant of $J$ and $\bJ$, $B = \frac{1}{2s} \bpd J =\frac{1}{2s} \pd \bJ $. The large $m$ value of $\lambda_{JJ\veps}$ and $\lambda_{J\bJ\veps}$ are in \eqref{eq:JJe_OPE}, and they satisfy \eqref{eq:BOPE}.

Let's now look at three point functions with one insertion of $A$. These will be more complicated, since $A$ is not a primary and logarithms appear. For convenience, we will use the short hand notation
\begin{equation}
 \llangle \calO_1(z_1) \calO_2(z_2) \calO_3(z_3) \rrangle = \frac{1}{z_{12}^{h_{123}} z_{23}^{h_{231}} z_{13}^{h_{132}}}\frac{1}{\bz_{12}^{\bar h_{123}} \bz_{23}^{\bar h_{231}} \bz_{13}^{\bar h_{132}}}
\end{equation}
to indicate the usual coordinate dependence one expects from an ordinary three point function of primaries, where $h_{ijk}=h_i+h_j-h_k$. In section \ref{sec:currents} we obtained two point functions by solving the associated conformal Ward identity \eqref{eq:wardId}; we repeat the same procedure for the three point functions and find
\beqa
\label{eq:AJeps}
\frac{\langle A(z_1)  J (z_2) \veps(z_3)  \rangle}{\llangle A(z_1)  J(z_2)  \veps (z_3) \rrangle} &= \lambda_{AJ\veps}-s\lambda_{J J \veps}\left( \frac{h_\veps}{h_\veps-1} \frac{z_{12}}{z_{23}}+\frac{\bz_{12}}{\bz_{23}}\right) + \frac{h_\veps\lambda_{JJ \veps}}{2s} \log \left| \frac{z_{12} z_{31}}{z_{23}}\right|^2\,, \\
\frac{\langle A(z_1) B(z_2)\veps(z_3) \rangle}{\llangle A(z_1) B(z_2) \veps(z_3) \rrangle} &=
\frac{\lambda_{AJ\veps}(h_\veps-1)}{2s}+\frac{h_\veps \lambda_{JJ\veps} }{4}\left(\frac{1}{s^2}- 2\frac{z_{12}}{z_{23}}-2\frac{\bz_{12}}{\bz_{23}}\right) +\frac{h_\veps(h_\veps-1)}{4s^2} \lambda_{J J \veps} \log \left| \frac{z_{12}z_{13}}{z_{23}} \right|^2\,.
\eeqa

Finally, the $\braket{AA\veps}$ three point function is 
\beqg
\frac{\langle A(z_1) A(z_2)\veps(z_3) \rangle}{\llangle A(z_1) A(z_2) \veps(z_3) \rrangle} =\lambda_{AA\veps}+l_0+l_1+l_2\\
l_2= \frac{1}{4s^2}\left(1-h_{\varepsilon }\right) h_{\varepsilon } \lambda _{\text{JJ$\varepsilon $}}\log \left|\frac{z_{12} z_{13}}{z_{23} }\right|^2 \log \left|\frac{z_{12} z_{23} }{z_{13} }\right|^2\\
l_1=\left[\frac{ \lambda_{AJ\veps}(1-h_\veps)}{s}-\lambda_{J J \veps}h_\veps \frac{s^2+1}{2s^2}+\frac 12 h_\veps \lambda_{JJ \veps} \left(1+\frac{z_{12}^2}{z_{13}z_{23}}+ \frac{\bz_{12}^2}{\bz_{13}\bz_{23}}\right)\right] \log\left|z_{12} \right|^2+\\
+\frac 12 h_\veps \lambda_{JJ\veps} \left(\frac{z_{13}}{z_{23}} -\frac{z_{23}}{z_{13}}+\frac{\bz_{13}}{\bz_{23}} -\frac{\bz_{23}}{\bz_{13}}\right) \log\left| \frac{z_{23}}{z_{13}} \right|^2\\
l_0=s(\lambda_{AJ\veps} +s \lambda_{J J \veps})\left(\frac{z_{13}}{z_{23}}+\frac{z_{23}}{z_{13}}+\frac{\bar{z}_{13}}{\bar{z}_{23}}+	\frac{\bar{z}_{23}}{\bar{z}_{13}}\right)+s^2 \lambda_{J J \veps} \frac{h_\veps}{h_\veps-1} \left(2+z_{12} \bar{z}_{12} \left(\frac{1}{z_{13} \bar{z}_{23}}+\frac{1}{z_{23} \bar{z}_{13}}\right)\right)
\eeqg
We remind the reader that $AA\veps$ OPE coefficients were not fixed by us uniquely from crossing; however, the leading behavior  was obtained by requiring a consistent $n\to2$ limit.
We always can shift $A\to A +\lambda B$, and this would change the OPE coefficients $\lambda_{AJ\veps}$ and $\lambda_{AA\veps}$. We keep $A$ so that its two point function is in the canonical form \eqref{eq:canonical}, and this makes the definition of both OPE coefficients unambiguous. 
The value of $s$ and the large $m$ behavior of $\lambda_{AJ\veps}$ can be found in \eqref{eq:s} and \eqref{eq:OPEA}. We also remind the reader that $h_\veps=1+O(1/m)$.

\small
\bibliography{../nBiblio}
\bibliographystyle{utphys}

\end{document}